\documentclass[12pt]{article}

\usepackage[utf8]{inputenc}
\usepackage[english]{babel}
\usepackage{amsmath}
\usepackage{amssymb}
\usepackage{bbm}
\usepackage{hyperref}
\usepackage{physics}
\usepackage{graphicx}
\usepackage{xcolor,colortbl}

\usepackage{cite}

\newcommand{\be}{\begin{equation}}
\newcommand{\ee}{\end{equation}}
\newcommand{\ba}{\begin{eqnarray}}
\newcommand{\ea}{\end{eqnarray}}
\newcommand{\bs}{\begin{subequations}}
\newcommand{\es}{\end{subequations}}
\newcommand{\no}{\nonumber\\}

\graphicspath{ {figures/} }

\begin{document}

\title{\LARGE Vacuum stability conditions for new $SU(2)$ multiplets}

\author{
  Andr\'e Milagre,$^{(1)}$\thanks{E-mail:
    \tt andre.milagre@tecnico.ulisboa.pt}
  \
  Darius~Jur\v{c}iukonis,$^{(2)}$\thanks{E-mail:
    \tt darius.jurciukonis@tfai.vu.lt}
  \
  and Lu\'\i s~Lavoura$^{(1)}$\thanks{E-mail:
    \tt balio@cftp.tecnico.ulisboa.pt}
  \\*[3mm]
  $^{(1)}\!$
  \small Universidade de Lisboa, Instituto Superior T\'ecnico, CFTP, \\
  \small Av.~Rovisco~Pais~1, 1049-001~Lisboa, Portugal
  \\*[2mm]
  $^{(2)}\!$
  \small Vilnius University, Institute of Theoretical Physics and Astronomy, \\
  \small Saul\.etekio~av.~3, Vilnius 10257, Lithuania
}

\maketitle

\begin{abstract}
  We consider the addition to the Standard Model
  of a scalar $SU(2)$ multiplet $\Delta_n$
  with dimension $n$ going from $1$ to $6$.
  The multiplet $\Delta_n$ is assumed to have
  null vacuum expectation value and an arbitrary (free) hypercharge.
  We determine the shape of the phase space
  for the new terms that appear in the scalar potential (SP);
  we observe in particular that,
  in the case of a 6-plet,
  the phase space is slightly concave along one of its boundaries.
  We determine the bounded-from-below and vacuum stability conditions
  on the SP for each value of $n$.
\end{abstract}

\section{Introduction}

The Higgs particle has been discovered at the Large Hadron Collider (LHC)
in 2012~\cite{Higgs1, Higgs2}.
Since then,
exploration of the interactions of that particle
has shown that they are quite close
to the predictions of the Standard Model (SM)~\cite{pdg}.
This either confirms that 
the breaking of the gauge symmetry of the SM
and the generation of the fermion masses are
effected solely by a single scalar doublet 
of $SU(2)$, 
or else it suggests the presence of an 
`alignment' mechanism \cite{Gunion:2002zf, Bernon:2015qea}
that allows a more complex scalar sector---such as in the 
two-Higgs-doublet model (2HDM)\cite{Branco:2011iw}---to mimic the SM 
predictions---despite the absence of a symmetry enforcing such alignment.
Moreover,
the almost-exact prediction of the SM
\be
\label{mwmz}
m_W = c_w m_Z
\ee
---where $m_W$ and $m_Z$ are the masses of the gauge bosons $W^\pm$ and $Z^0$,
respectively,
and $c_w$ is the cosine of the weak mixing angle---strongly
suggests that only $SU(2)$ doublets,
and possibly also singlets,
have vacuum expectation values (VEVs)~\cite{albergaria}.
So,
the scalar sector of any extension of the SM
is already today rather strongly constrained.
On the other hand,
there is no reason---but for Occam's razor---why the scalar sector
of a spontaneously broken $SU(2) \times U(1)$ gauge theory
should consist of only one $SU(2)$ doublet.
One can entertain the speculation that larger $SU(2)$ multiplets exist---even
if they have zero VEV because of Eq.~\eqref{mwmz},
and even if they are so large that they do not couple to the known fermions.
In particular,
large extra $SU(2)$ multiplets may be useful,
even if they have no VEV,
to alter Eq.~\eqref{mwmz} through the radiative (`oblique') corrections
that they produce,
if that Eq.~\eqref{mwmz} is observed to be slightly off the mark~\cite{chinese}.

In this paper we consider the possibility that the scalar sector
of an $SU(2) \times U(1)$ gauge theory consists of
one hypercharge-$1/2$ $SU(2)$ doublet $\Phi$,
which has VEV,
and another $SU(2)$ multiplet $\Delta_n$,
with weak isospin $J$ and $n = 2 J + 1$ components,
that has null VEV and a free (arbitrary) hypercharge.
The latter assumption means that
the theory has a global $U(1)$ symmetry
$\Delta_n \to \Delta_n \exp \left( i \vartheta \right)$;
since $\Delta_n$ has zero VEV,
that symmetry remains unbroken, and no Goldstone boson 
arises from it.\footnote{Two other recent papers that consider
the addition to the SM of $SU(2)$ scalar multiplets with dimension
up to $n=5$ are~\cite{Giarnetti:2023osf,Giarnetti:2023dcr}. 
Note, however, that in those papers, those multiplets 
are supposed to acquire VEVs, contrary to what we assume here.}

A difficult problem that one faces when one considers any extension
of the scalar sector of the SM is to find the bounded-from-below (BFB)
conditions\footnote{Some people call `vacuum stability conditions'
to what we call `BFB conditions'.
We reserve the former term
to the conditions stemming from another rationale (see below).}
on the scalar potential (SP) $V$.
This is a mathematical problem that is very easy to state---what are
the conditions on the coupling constants
of the quartic part $V_4$ of the SP,\footnote{We assume
that the SP is renormalizable.}
such that $V_4$ can be negative for \emph{no} configuration
of the scalar fields\footnote{If $V_4$ is negative
for some values of the scalar fields,
then by multiplying all the fields by an ever larger positive real number 
$\kappa$
one makes $V_4 \to \kappa^4 V_4$,
which is ever more negative,
and this means that $V$ is not BFB and therefore has no minimum,
\textit{i.e.}\ the theory lacks vacuum state.}---but
surprisingly difficult to solve
even for modest extensions of the SM.
The problem has only been solved for the two-Higgs-doublet
model~\cite{Maniatis, Ivanov},
for some constrained forms of the three-Higgs-doublet
model~\cite{Faro, Ivanov2020, Buskin, Boto, Carrolo},
for a few models with $SU(2)$ triplets~\cite{arhrib, fonseca, Moultaka},
and for a few other rather simple models~\cite{Dirgantara,kannikerecent,our4}.
Additionally,
there are in the literature BFB conditions 
for the $SU(3) \times U(1)$ electroweak model~\cite{constantini}
and for models with coloured scalars~\cite{Heikinheimo}.

There are papers on general methods
for deriving BFB conditions~\cite{copositive, Kannike2016, muhlleitner}.
Among other methods,
geometric and group-theoretic approaches
may sometimes be utilized. 
These include analyzing the potential
on the \emph{orbit space}~\cite{Abud1, Kim1, Sartori2003,
  Ivanov2006, Ivanov2010-I, degee}, 
applying \emph{stratification theory}
(the classification of extrema by symmetry patterns)~\cite{Michel,
  Abud2, Ivanov2010-II, Ivanov2020}, 
and using \emph{boundary conditions and convexity
(copositivity) criteria}~\cite{copositive, Barroso, Heikinheimo,
  Kannike2016}.
The orbit space approach
seeks to find
a coordinate system tailored to the symmetries of the potential,
often transforming
the problem of the minimization of $V_4$
into one of analyzing geometric shapes (cones, polyhedra, etc.)
within that
space. 
Due to its versatility,
many modern analyses of multi-scalar potentials---including those
employing the $P$-matrix
formalism~\cite{kannikerecent,Dirgantara}---either
explicitly or implicitly utilize orbit-space reasoning.

In this paper, we analyze the gauge orbit space
(which we call the \textit{phase space}) to derive
BFB constraints for two $SU(2)$ scalar multiplets
$\Phi$ and $\Delta_n$.
When the potential is linear in the phase space variables,
it is sufficient to consider the convex hull of the orbit space
to determine the BFB conditions~\cite{Heikinheimo}---concave stretches of
boundary do not matter.
We note, however,
that for two multiplets the potential has the form in Eq.~\eqref{matrix_V}
below,
and then one needs,
at least in principle,\footnote{We do not do that explicitly
in the case $n=6$,
because the concavity existing in that case is extremely slight.}
to take into account the concave stretches of the boundary in detail.

One further aim of this paper is to fill a gap
in our understanding of vacuum stability in multi-scalar models,
at least partially.
(We call `vacuum stability conditions'
to the conditions for the desired minimum of the potential
to be its absolute \emph{global} minimum,
and not just a local minimum.
Note that,
contrary to the BFB conditions,
that only affect the quartic part $V_4$ of $V$,
the vacuum stability conditions affect the whole $V$
including its quadratic part.)
Building on the results of Refs.~\cite{our, Milagre:2024nvy},
we derive analytical vacuum stability constraints
for the scalar potentials of SM extensions
through scalar multiplets up to $n=6$.
The detailed expressions and explanations provided in the paper 
may be useful for readers attempting to apply the methodology 
to other models.

The plan of this paper is as follows.
After writing down in Section~\ref{sec:SP}
the most general renormalizable SP for our model,
we delimit in Section~\ref{sec:OS}
the extent of the freedom of that SP---which is smaller than
its many scalar fields suggest,
because the SP is renormalizable,
\textit{i.e.}\ it contains no higher powers of the fields than four.
That allows us to derive,
in Section~\ref{sec:BFB},
necessary and sufficient (n\&s) conditions for $V_4$ to be BFB
when $n \le 5$---and almost n\&s conditions when $n=6$.
In Section~\ref{sec:vacuumstability}
we deal on the vacuum stability conditions
and in Section~\ref{sec:conclusions} we summarize our achievements.
Appendix~\ref{app:previous} collects useful results
from a previous paper of two of us~\cite{our},
Appendix~\ref{app:phaseSpace}
considers \textit{Ans\"atze} for the fields in the cases
$n = 5$ and $n = 6$,
and Appendix~\ref{app:math} solves a technical mathematical problem
that often arises in the main body of the paper.

\section{The scalar potential}
\label{sec:SP}

We write the Higgs doublet of the SM as
\be
\Phi = \left( \begin{array}{c} a \\ b \end{array} \right),
\ee
where $a$ and $b$ are complex scalar fields.
Then,
\be
\widetilde \Phi = \left( \begin{array}{c} b^\ast \\ - a^\ast \end{array} \right)
\ee
is also an $SU(2)$ doublet.
Let $I_z$ be the third component of isospin of the generic field $z$.
One has $I_a = - I_b = 1/2$.
The SP of the SM is
\be
V_ \mathrm{SM} = \mu_1^2 F_1 + \frac{\lambda_1}{2}\, F_1^2,
\ee
where
\be
F_1 = \left| a \right|^2 + \left| b \right|^2 \equiv A + B
\label{F1}
\ee
is $SU(2)$-invariant.
In general, we denote the squared modulus of the generic field $z$
by the corresponding capital letter $Z$,
\textit{i.e.}\ $Z \equiv \left| z \right|^2$.

In the models that we consider in this paper
there is just one scalar $SU(2)$ multiplet beyond $\Phi$;
we call that extra multiplet $\Delta_n$,
where the integer $n = 1, \ldots, 6$ denotes the dimension 
of the irreducible representation of $SU(2)$ embodied by $\Delta_n$.
Of course $n = 2 J + 1$,
where $J$ is the weak isospin of $\Delta_n$.
If $n \ge 3$ and an electrically neutral component of $\Delta_n$
has vacuum expectation value (VEV),
then the gauge-boson masses do not obey Eq.~\eqref{mwmz}.
Indeed~\cite{albergaria},
\bs
\label{mWmZ2}
\ba
m_W^2 &=& g^2 \sum_{\varphi_{JY}} \left| v_{JY} \right|^2
\left( J^2 - Y^2 + J \right),
\\
m_Z^2 &=& \frac{g^2}{\cos^2 {\theta_w}}\,
  \sum_{\varphi_{JY}} \left| v_{JY} \right|^2 \left( 2 Y^2 \right),
\ea
\es
where the sum is performed over all neutral fields 
$\varphi_{JY}$ with isospin $J$, hypercharge $Y$,
and VEV $v_{JY}$.
In Eqs.~\eqref{mWmZ2},
$g$ is the $SU(2)$ gauge coupling constant.

In order to keep the relation~\eqref{mwmz} valid,
we assume that all the neutral fields except $b$
have null VEV.
Partly in order to guarantee this,
we also assume that the models enjoy $U(1)$ symmetries
$\Delta_n \to \exp \left( i \vartheta \right) \Delta_n$;
those symmetries prevent couplings
either of the form $\Phi^2 \Delta_n$ or of the form $\Phi^3 \Delta_n$,
that might induce a VEV for a component of $\Delta_n$.\footnote{Such couplings
would anyway be forbidden by the $SU(2)$ gauge symmetry,
together with renormalizability,
for $n > 4$.}
With this additional symmetry,
the SPs become simpler
and are given by a general form explained in Ref.~\cite{our}
and given in detail in Appendix~\ref{app:previous}.
The most important takeaway is that
\bs
\label{7}
\ba
\textup{when $n=1$,}
&&
V = V_\mathrm{SM} + \mu_2^2 F_2 + \frac{\lambda_2}{2}\, F_2^2 
+ \lambda_3 F_1 F_2; \label{7a}
\\
\textup{when $n=2$,}
&&
V = V_\mathrm{SM} + \mu_2^2 F_2 + \frac{\lambda_2}{2}\, F_2^2 
+ \lambda_3 F_1 F_2 + \lambda_4 F_4; \label{7b}
\\
\textup{when either $n=3$ or $n=4$,}
&&
V = V_\mathrm{SM} + \mu_2^2 F_2 + \frac{\lambda_2}{2}\, F_2^2 
+ \lambda_3 F_1 F_2
+ \lambda_4 F_4 + \lambda_5 F_5; \label{7c}
\\
\textup{when either $n=5$ or $n=6$,}
&&
V = V_\mathrm{SM} + \mu_2^2 F_2 + \frac{\lambda_2}{2}\, F_2^2 
+ \lambda_3 F_1 F_2
+ \sum_{k=4}^6 \lambda_k F_k. \label{7d}
\ea
\es
In Eqs.~\eqref{7},
the $F_k$ are $SU(2)$-invariant polynomials
of the fields of $\Phi$ and of $\Delta_n$;
their functional forms depend on the dimension of $\Delta_n$.

\section{Phase spaces}
\label{sec:OS}

The vacuum structure of a scalar potential may be analyzed geometrically
by studying the space spanned by the invariants of the theory,
which we call the \emph{phase space}\footnote{Other authors 
call it
the `orbit space'~\cite{Abud1, Kim1, Sartori2003,
Ivanov2006, Ivanov2010-I, degee}.
}.
In particular, the conditions for boundedness from below and vacuum stability
may be derived from the shape of the phase space.
However,
for arbitrary field configurations,
the invariants $F_i$ may become unbounded,
leading to a loss of information about the boundary of the phase space. 
In order to avoid this,
we introduce the following set
of dimensionless $SU(2)$-invariants\footnote{We employ the same notation
as Ref.~\cite{our},
except that our $\delta$ is rescaled by a factor of $J/2$.
Therefore,
when comparing the results between the two papers
one should take into account that
\be
\delta_{\textup{here}} = \frac{J}{2}\, \delta_{\textup{Ref.\cite{our}}}.
\ee
}
\be
r\equiv \frac{F_1}{F_2}, 
\quad\quad \gamma_5 \equiv \frac{F_5}{F_2^2},
\quad\quad \gamma_6 \equiv \frac{F_6}{F_2^2}, 
\quad\quad \delta \equiv \frac{F_4}{F_1 F_2},
\label{invariants}
\ee
enabling us to rewrite the SP as
\be
V =
\mu_1^2 \,F_1 + \mu_2^2 \,F_2 
+  \left[ \frac{\lambda_1}{2}\, r^2
  + \Pi\left(\delta \right) r
  + \frac{\Xi\left(\gamma_5, \gamma_6 \right)}{2} \right] F_2^2,
\label{THESP}
\ee
where
\bs
\label{PiXi}
\ba
\Pi \left( \delta \right) &\equiv& \lambda_3 + \lambda_4\,\delta, \label{Pi}
\\
\Xi \left( \gamma_5, \gamma_6 \right) &\equiv& \lambda_2
+ 2 \lambda_5 \gamma_5 + 2 \lambda_6 \gamma_6. \label{Xi}
\ea
\es
In the following subsections,
we derive the shape of the space spanned
by the $SU(2)$-invariants $\gamma_5$,
$\gamma_6$,
and $\delta$.
We will refer to this space as \textit{the} phase space,
even though,
in fact,
it is just a subspace of the true phase space that also 
includes the unbounded invariants $F_1$ and $F_2$.
Since $\delta=\gamma_5=\gamma_6=0$ for the case $n=1$,
its corresponding phase space is zero-dimensional,
and we start at $n=2$.

\subsection{$n=2$}

In the case $n=2$, \textit{i.e.} $J=1/2$,
both $\gamma_5$ and $\gamma_6$ are zero,
hence the phase space is spanned by a single dimensionless parameter $\delta$. 
From its definition in Eq.~\eqref{invariants} we find
\bs
\ba
\delta &=& \frac{1}{4}
- \frac{\left| a d - b c \right|^2}{2
  \left( A + B \right) \left( C + D \right)} \label{12a}
\\ &=& - \frac{1}{4}
+ \frac{\left| a c^\ast + b d^\ast \right|^2}{2
  \left( A + B \right) \left( C + D \right)}. \label{12b}
\ea
\es
Since the second terms in the right-hand sides of Eqs.~\eqref{12a}
and~\eqref{12b} are non-negative,
\be
\left| \delta \right| \le \frac{1}{4}.
\label{hhhhhh}
\ee

\subsection{$n=3$}

In the case $n=3$, \textit{i.e.} $J=1$,
the phase space has two dimensionless parameters $\delta$ and $\gamma_5$.
One may show that~\cite{our}
\be
1 - 4 \delta^2 - 3 \gamma_5 \ge 0,
\label{14}
\ee
which simultaneously bounds $\delta$ and $\gamma_5$;
the latter moreover is,
by definition,
non-negative~\cite{our}.
Therefore,
\bs
\label{15}
\ba
0 \ \le \ \left| \delta \right| &\le& \frac{\sqrt{1-3 \gamma_5}}{2},
\label{3vjofof}
\\
0 \ \le \ \gamma_5 &\le& \frac{1}{3}. \label{eq:eeeeeww}
\ea
\es
The parameter $\gamma_5$ is a function of the three fields $c$,
$d$,
and $e$ through
\be
\gamma_5 = \frac{\left| 2 c e - d^2 \right|^2}{3
  \left( C + D + E \right)^2}.
\ee
The upper bound~\eqref{eq:eeeeeww} on $\gamma_5$ is saturated,
for instance,
if $c=e=0$,
or else if $d=0$ and $c=e$.
This case was first rigorously treated in Ref.~\cite{fonseca},
after an initial attempt in Ref.~\cite{arhrib}.

\subsection{$n=4$}

In the case $n=4$, \textit{i.e.} $J=3/2$,
the phase space once again has two parameters $\delta$ and $\gamma_5$.
One may show that they are simultaneously bounded by the condition~\cite{our}
\be
9 - 16\,\delta^2 - 20\gamma_5 \ge 0,
\ee
which substitutes Eq.~\eqref{14} of case $n=3$.
This means that now
\bs
\label{17}
\ba
0 \ \le \ \left| \delta \right| &\le& \frac{\sqrt{9 - 20 \gamma_5}}{4},
\label{delta44444}
\\
0 \ \le \ \gamma_5 &\le& \frac{9}{20},\label{eqggtsda}
\ea
\es
instead of Eqs.~\eqref{15}.
The parameter $\gamma_5$ is a function of the four fields $c, \ldots, f$ through
\be
\gamma_5 = \frac{2 \left| \sqrt{3}\, c e - d^2 \right|^2
  + 2 \left| \sqrt{3}\, d f - e^2 \right|^2
  + \left| 3 c f - d e \right|^2}{5 \left( C + D + E + F\right)^2}.
\ee
The upper bound~\eqref{eqggtsda} on $\gamma_5$ is saturated, for instance, when $d=e=0$ and $c=f$.

\subsection{$n=5$}

In the case $n=5$, \textit{i.e.} $J=2$,
the phase space has three dimensionless parameters $\delta$,
$\gamma_5$,
and $\gamma_6$.
The condition~\cite{our}
\be
4 - 4 \delta^2 - 7 \gamma_5 - 10 \gamma_6 \ge 0
\label{hvjvpp}
\ee
simultaneously bounds $\delta$,
$\gamma_5$,
and $\gamma_6$.
The upper bound on the magnitude of $\delta$ now reads
\be
\left|\delta\right| \le \frac{\sqrt{4 - 7\gamma_5 - 10\gamma_6}}{2}.
\label{delta555555}
\ee

\begin{figure}[h!]
\centering
\includegraphics[width=0.52\textwidth]{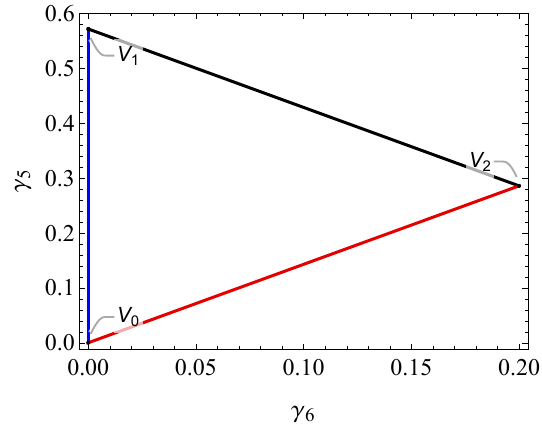}
\caption{Boundaries of the phase space for the case $n=5$,
plotted in the $\left( \gamma_6, \gamma_5 \right)$ plane.
The blue straight line is given by Eq.~\eqref{side1},
the red straight line is given by Eq.~\eqref{side2},
and the black straight line is given by Eq.~\eqref{side3}.
The vertices are given by Eqs.~\eqref{vertices}.}
\label{fig:phase_5}
\end{figure}

The invariants $\gamma_5$ and $\gamma_6$ may be written in terms of $F_2$,
$F_5$,
and $F_6$,
which in turn are given in terms of the five fields $c, \ldots g$
by Eqs.~\eqref{A17}.
Since $F_2$,
$F_5$,
and $F_6$ are non-negative,
$\gamma_5$ and $\gamma_6$ are non-negative too.
In order to gain a grasp on how large $\gamma_5$ and $\gamma_6$ may be,
we have considered in section~\ref{app:phaseSpace5}
of Appendix~\ref{app:phaseSpace}
three \textit{Ans\"atze} for $c, \ldots, g$.
With these three \textit{Ans\"atze} we have constructed a triangle in the
$\left( \gamma_6, \gamma_5 \right)$ plane,
which we depict in Fig.~\ref{fig:phase_5}.
The sides of that triangle are
given by Eqs.~\eqref{side1},
\eqref{side2},
and~\eqref{side3}, and are plotted in
blue, red, and black, respectively.
The vertices of the triangle are
\bs
\label{vertices}
\ba
V_0 &=& \left( 0,\ 0 \right),
\\
V_1 &=& \left( 0,\ \frac{4}{7} \right),
\\
V_2 &=& \left( \frac{1}{5},\ \frac{2}{7} \right).
\ea
\es

Numerically generating random complex values for the five fields $c, \ldots, g$
and therefrom computing $\gamma_5$ and $\gamma_6$
by means of Eqs.~\eqref{invariants} and~\eqref{A17},
one finds that all the $\left( \gamma_6, \gamma_5 \right)$ thus obtained
are inside the above-mentioned triangle.
So,
\textbf{the range of $\left( \gamma_6, \gamma_5 \right)$ is the triangle
with vertices $V_0$,
$V_1$,
and $V_2$ in Eqs.~\eqref{vertices}}.

\subsection{$n=6$}

In the case $n=6$, \textit{i.e.} $J=5/2$,
the phase space once again has parameters $\delta$,
$\gamma_5$,
and $\gamma_6$.
One may show that these $SU(2)$-invariants
are simultaneously bounded by the condition~\cite{our}
\be
25 - 16\delta^2 - 36 \gamma_5 - 56 \gamma_6 \ge 0,
\label{22}
\ee
instead of by inequality~\eqref{hvjvpp}.
In turn,
the upper-bound on the magnitude of $\delta$ now reads
\be
\left| \delta \right| \le \frac{\sqrt{25 -36\gamma_5 -56\gamma_6}}{4}
\label{delta666666}
\ee
instead of inequality~\eqref{delta555555}.

\begin{figure}[h]
\centering
\includegraphics[width=0.95\textwidth]{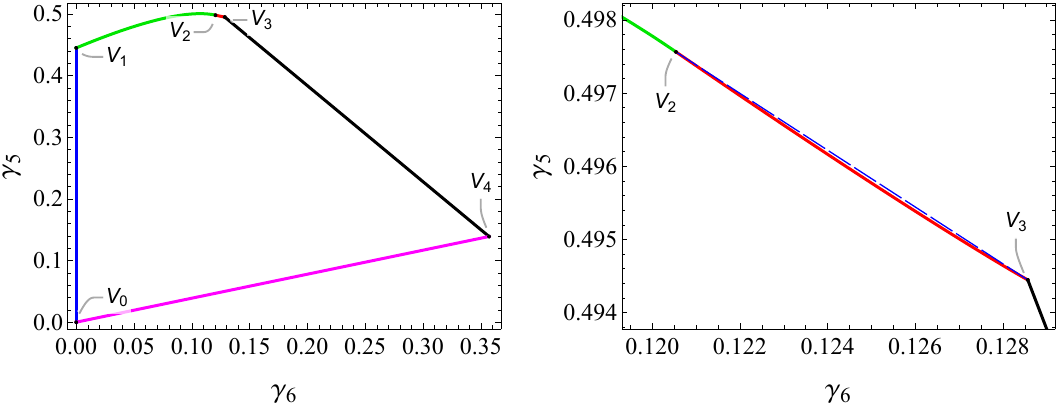}
\caption{
Left: Boundaries of the phase space for the case $n=6$,
plotted in the $\left( \gamma_6, \gamma_5 \right)$ plane.
The points $V_0, \ldots, V_4$ are given by Eqs.~\eqref{vert_6}.
The blue straight line is given by Eq.~\eqref{943oo};
the green curve is given by Eq.~\eqref{3nvjfod};
the red curve is given in parametric form by Eqs.~\eqref{parametric};
the black straight line is given by Eq.~\eqref{jgf000r};
the magenta straight line is given by Eq.~\eqref{prop}.
Right: zoomed-in view of the region of points $V_2$ and $V_3$.
The dashed blue straight line is given by Eq.~\eqref{parametric_approx}.
}
\label{fig:phase_6}
\end{figure}

The invariants $\gamma_5$ and $\gamma_6$ 
may be written in terms of the six fields $c, \ldots h$
through Eqs.~\eqref{invariants} and~\eqref{A20};
they are non-negative because $F_2$,
$F_5$,
and $F_6$ are also non-negative.
In order to find out the range spanned by $\gamma_5$ and $\gamma_6$,
we have firstly considered five \textit{Ans\"atze}\/ for the fields,
given in detail in section~\ref{app:phaseSpace6}
of Appendix~\ref{app:phaseSpace}.
With these five \textit{Ans\"atze} we have formed the boundary of a domain
in the $\left( \gamma_6, \gamma_5 \right)$ plane.
That domain is depicted in the left panel of Fig.~\ref{fig:phase_6}
and has five vertices\footnote{Vertex $V_2$ is not really a vertex;
it is just the point
where the curves~\eqref{3nvjfod} and~\eqref{parametric} meet.
At that point,
the two curves have the same slopes but different second derivatives.
Indeed, the curve~\eqref{3nvjfod} is a \emph{convex} boundary
of the $\left( \gamma_6, \gamma_5 \right)$ domain,
while the curve~\eqref{parametric} is a \emph{concave} boundary.}:
\bs
\label{vert_6}
\ba
V_0 &=& \left( 0,\ 0 \right);
\\
V_1 &=& \left( 0,\ \frac{4}{9} \right);
\\
V_2 &=& \left( \varrho,\ \varsigma \right);
\\
V_3 &=& \left( \frac{9}{70},\ \frac{89}{180} \right);
\\
V_4 &=& \left( \frac{5}{14},\ \frac{5}{36} \right).
\ea
\es
To better illustrate the slight concavity of the boundary 
connecting vertices $V_2$ and $V_3$, 
the right panel
of Fig.~\ref{fig:phase_6}
shows a zoomed-in view of this region. 
The blue dashed line represents the approximation of Eqs.~\eqref{parametric}
by the straight line in Eq.~\eqref{parametric_approx}.

Numerically generating random complex values for the six fields
$c, \ldots, h$ and therefrom computing $\gamma_5$ and $\gamma_6$
by means of Eqs.~\eqref{A20} and~\eqref{invariants},
one finds that all the points $\left( \gamma_6, \gamma_5 \right)$ thus generated
are inside the above-mentioned domain,
and indeed fill it completely.
So,
\textbf{the range of $\left( \gamma_6, \gamma_5 \right)$ is the area
  bounded by the lines~\eqref{943oo},
  \eqref{3nvjfod},
  \eqref{parametric},
  \eqref{jgf000r},
  and~\eqref{prop}}.
This area is very slightly \emph{concave} at one of its borders,
\textit{viz.}\ at the line~\eqref{parametric}.\footnote{The 
possibility that the phase space
is concave at some of its boundaries was already made clear
in Fig.~2 of Ref.~\cite{Heikinheimo}.
That was in a model with a four-dimensional
phase space; here we find the same feature in a simpler model
with a two-dimensional phase space.}

\section{Boundedness from below}
\label{sec:BFB}

The SP in Eq.~\eqref{THESP} is BFB if
its quartic part ($V_4$) is non-negative for every possible field configuration,
\textit{i.e.} for every point in phase space.
Following the method outlined in Refs.~\cite{our, copositive},
we rewrite the quartic part of the SP as
\be
\frac{V_4}{F_2^2} = \frac{1}{2}\,
\left( \begin{array}{cc} r, & 1 \end{array} \right)
\left( \begin{array}{cc} \lambda_1 & \Pi\left(\delta \right)\\*[1mm]
  \Pi\left(\delta \right) & \Xi\left(\gamma_5, \gamma_6 \right)
\end{array} \right)
\left( \begin{array}{c} r \\ 1 \end{array} \right),
\label{matrix_V}
\ee
where $\Pi \left( \delta \right)$
and $\Xi\left(\gamma_5, \gamma_6 \right)$
have been defined in Eqs.~\eqref{PiXi}.
Since $r$ is strictly positive,
the SP in Eq.~\eqref{THESP} is BFB
if and only if the $2 \times 2$ matrix in Eq.~\eqref{matrix_V}
is copositive~\cite{copositive}.
Therefore,
we must ensure that the following conditions
hold for every $\delta$,
$\gamma_5$,
and $\gamma_6$:
\bs
\label{BFBall}
\ba
\lambda_1 &\ge & 0, \label{BFB1} \\
\Xi\left(\gamma_5, \gamma_6 \right) &\ge & 0, \label{BFB2} \\
\widetilde{\Gamma}\left(\delta, \gamma_5, \gamma_6 \right) \equiv
\Pi \left( \delta \right)
+ \sqrt{ \lambda_1 \,\Xi\left(\gamma_5, \gamma_6 \right)} &\ge& 0.
\label{BFB3}
\ea
\es
As stated in Ref.~\cite{fonseca},
it is \emph{necessary} that $\Xi\left(\gamma_5, \gamma_6 \right)$
and $\widetilde{\Gamma}\left(\delta, \gamma_5, \gamma_6 \right)$ are 
non-negative
everywhere on the boundary of the phase space
for conditions \eqref{BFBall} to hold.
In addition,
if the absolute minima of those functions lie inside 
the phase space,
then one must require them to be non-negative.
Regarding $\Xi\left(\gamma_5, \gamma_6 \right)$ given in Eq.~\eqref{Xi},
we note that it is monotonic in both $\gamma_5$ and $\gamma_6$,
so its minimum necessarily lies at the boundary of the phase space.
On the other hand,
$\widetilde{\Gamma}\left(\delta, \gamma_5, \gamma_6 \right)$
is monotonic in $\delta$,
hence its minimum is attained at the boundary of the phase space in the 
$\delta$ direction.
We affect the minimization with 
respect to $\delta$ firstly and fix
\be
\delta = -\frac{\lambda_4}{\left| \lambda_4 \right|}\,
\Sigma\left(\gamma_5, \gamma_6 \right),
\ee
where,
by using Eqs.~\eqref{hhhhhh},
\eqref{3vjofof},
\eqref{delta44444},
\eqref{delta555555},
and \eqref{delta666666} we have
\bs
\ba
&&\Sigma\left(\gamma_5, \gamma_6 \right) 
= 0 \quad \textup{when $n=1$},\\
&&\Sigma\left(\gamma_5, \gamma_6 \right) 
= \frac{1}{4} \quad \textup{when $n=2$},\\
&&\Sigma\left(\gamma_5, \gamma_6 \right) 
= \frac{\sqrt{1-3 \gamma_5}}{2} 
\quad \textup{when $n=3$},\\
&&\Sigma\left(\gamma_5, \gamma_6 \right) 
= \frac{\sqrt{9-20 \gamma_5}}{4} 
\quad \textup{when $n=4$},\\
&&\Sigma\left(\gamma_5, \gamma_6 \right) 
= \frac{\sqrt{4-7 \gamma_5- 10\gamma_6}}{2} 
\quad \textup{when $n=5$},\\
&&\Sigma\left(\gamma_5, \gamma_6 \right) 
= \frac{\sqrt{25 - 36\gamma_5 - 56\gamma_6}}{4}
\quad \textup{when $n=6$}.
\ea
\es
Then, the functions that we ought to minimize
are $\Xi\left(\gamma_5, \gamma_6 \right)$ of Eq.~\eqref{Xi} and
\be
\label{funcMIN}
\Gamma\left(\gamma_5, \gamma_6 \right) =
\lambda_3 - \left|\lambda_4\right|\,
\Sigma\left(\gamma_5, \gamma_6 \right)
+\sqrt{ \lambda_1 \,\Xi\left(\gamma_5, \gamma_6 \right)}.
\ee
In the following sections, we deduce the necessary and sufficient
conditions for the SP to be BFB when $n=1,\ldots,6$.
We note that the conditions presented in Ref.~\cite{our}
for the cases $n=1, \ldots, 4$ are necessary and sufficient,
but those for the cases $n=5$ and $n=6$ are just \emph{necessary}.
We chose to write them for all $n$ for the sake of completeness.

\subsection{The cases $n=1, \ldots, 4$}

\paragraph{$n=1$:} As previously noted,
the phase space for the case $n=1$ is zero-dimensional.
Setting $\delta = \gamma_5 = \gamma_6 = 0$ in Eqs.~\eqref{BFBall},
those BFB conditions become
\bs
\ba
\lambda_1 &\geq& 0,
\\
\lambda_2 &\geq& 0,
\\
\lambda_3 + \sqrt{\lambda_1 \lambda_2} &\geq& 0.
\ea
\es
These conditions are in agreement with those derived
for the complex singlet extension of the 
SM, for instance in Ref.~\cite{Barger:2008jx}.

\paragraph{$n=2$:} In the case $n=2$ we have $\gamma_5 = \gamma_6 = 0$.
Consequently,
%
\be
\Xi \left( \gamma_5, \gamma_6 \right) = \lambda_2, 
\quad
\Gamma \left( \gamma_5, \gamma_6 \right)
= \lambda_3 - \frac{\left|\lambda_4\right|}{4}
+ \sqrt{ \lambda_1 \lambda_2}.
\ee
Enforcing the conditions~\eqref{BFBall},
the BFB conditions read
\bs
\label{cmpdpdp}
\ba
\lambda_1 &\geq& 0,
\\
\lambda_2 &\geq& 0,
\\
\lambda_3 - \frac{\left|\lambda_4\right|}{4}
+\sqrt{ \lambda_1 \lambda_2} &\geq& 0.
\ea
\es
If one chooses to use the usual notation
for the case of the $U(1)$-symmetric 2HDM,
given in Eqs.~\eqref{theequivalence},
then conditions~\eqref{cmpdpdp} read
\bs
\ba
\bar \lambda_1 &\geq& 0,
\\
\bar \lambda_2 &\geq& 0,
\\
\bar \lambda_3 + \frac{\bar \lambda_4}{2}
- \frac{\left| \bar \lambda_4\right|}{2}
+ \sqrt{ \bar \lambda_1 \bar \lambda_2} &\geq& 0.
\label{preovc}
\ea
\es
Condition~\eqref{preovc} is equivalent to
\bs
\ba
\bar \lambda_3
+ \sqrt{ \bar \lambda_1 \bar \lambda_2} &\geq& 0,
\\
\bar \lambda_3 + \bar \lambda_4
+ \sqrt{ \bar \lambda_1 \bar \lambda_2} &\geq& 0.
\ea
\es
Therefore,
our conditions~\eqref{cmpdpdp} are equivalent to
\bs
\ba
\bar \lambda_1 &\geq& 0,
\\
\bar \lambda_2 &\geq& 0,
\\
\bar \lambda_3 &\geq& 
- \sqrt{ \bar \lambda_1 \bar \lambda_2},
\\
\bar \lambda_3 + \bar \lambda_4 &\geq& 
- \sqrt{ \bar \lambda_1 \bar \lambda_2},
\ea
\es
which are the textbook BFB conditions for the 
$U(1)$-symmetric 2HDM \cite{Branco:2011iw, Ginzburg:2010wa}.

\paragraph{$n=3$:} In the case $n=3$ only $\gamma_6$ is $0$,
so we need to minimize the functions
\bs
\ba
\Xi\left(\gamma_5\right) &=& \lambda_2 + 2\lambda_5 \gamma_5,
\\
\Gamma\left(\gamma_5\right) &=&
\lambda_3 - \frac{\left|\lambda_4\right|}{2}\, \sqrt{1-3 \gamma_5} 
+\sqrt{ \lambda_1}\, \sqrt{\lambda_2 + 2\lambda_5 \gamma_5}.
\ea
\es
As stated before, $\Xi\left(\gamma_5\right)$ is monotonic,
so it is non-negative everywhere in phase space
if it is non-negative at its end-points.
Explicitly, we require
\ba
\label{bfbg_3}
\left\{
\begin{array}{l}
  \displaystyle{\Xi(0)} \geq 0,
  \\*[2mm]
  \displaystyle{\Xi\left( \frac{1}{3} \right)} \geq 0,
\end{array}
\right.
& \Leftrightarrow &
\left\{
\begin{array}{l}
  \displaystyle{\lambda_2} \geq 0,
  \\*[2mm]
  \displaystyle{\lambda_2 + \frac{2}{3}\, \lambda_5} \geq 0.
\end{array}
\right.
\ea
Furthermore, $\Gamma\left(\gamma_5\right)$ is of the form of 
the function analyzed in Appendix~\ref{app:math},
with
\begin{equation}
	k = \frac{1}{2},\quad
	p = 1,\quad
	s = 3,\quad
	w = 2\lambda_5,\quad
	v = \lambda_2,\quad
	\alpha = 0,\quad
	\beta = \frac{1}{3}.
\end{equation}
For it to be non-negative everywhere inside the $\gamma_5$ domain
$\left[ \alpha, \beta \right]$ it is necessary to require
\bs
\label{condiiiiabd}
\ba
\lambda_3 - \frac{\left| \lambda_4 \right|}{2}
+ \sqrt{\lambda_1 \lambda_2} &\geq& 0,
\\
\lambda_3 + \sqrt{\lambda_1 
\left(\lambda_2 + \frac{2}{3}\lambda_5\right)} &\geq& 0.
\ea
\es
Furthermore, assuming conditions~\eqref{bfbg_3} to hold,
it is sufficient to \emph{exclude} the situation where
\bs
\label{condi_exclude}
\ba
\lambda_5 &<& -\frac{3 \lambda_4^2}{8 \lambda_1},
\\
\lambda_5 &<&
-\frac{3}{4}\, \sqrt{\frac{\lambda_2}{\lambda_1}}
\left|  \lambda_4 \right|,
\\
\lambda_3 &<& -
\sqrt{\frac{\left(2\lambda_5 + 3\lambda_2 \right)
\left(8\lambda_5 \lambda_1 + 3 \lambda_4^2 \right)}{24\lambda_5}}.
\ea
\es

Despite their differing appearance, the n\&s conditions 
presented in Eqs.~\eqref{bfbg_3}, \eqref{condiiiiabd}, 
and \eqref{condi_exclude} are Boolean-equivalent to 
those originally derived in~\cite{fonseca} and 
later confirmed in~\cite{our,Moultaka:2020dmb}.
We have confirmed this equivalence both numerically and algebraically.

\paragraph{$n=4$:} In the case $n=4$,
$\gamma_6 = 0$ and the functions that we need to minimize take the form
\bs
\ba
\Xi\left(\gamma_5\right) &=&
\lambda_2 + 2\lambda_5 \gamma_5,\\
\Gamma\left(\gamma_5\right) &=&
\lambda_3 - \frac{\left|\lambda_4\right|}{4}\, \sqrt{9-20 \gamma_5} 
+\sqrt{ \lambda_1}\, \sqrt{\lambda_2 + 2\lambda_5 \gamma_5}.
\ea
\es
Again,
$\Xi\left(\gamma_5\right)$ is monotonic in $\gamma_5$,
implying that it is non-negative everywhere for
$\gamma_5 \in \left[ 0, 9/20 \right]$ if and only if
it is non-negative both for $\gamma_5 = 0$ and for $\gamma_5 = 9/20$,
\textit{i.e.}
\bs
\ba
\lambda_2 &\geq& 0,
\\
\lambda_2 + \frac{9 \lambda_5}{10} &\geq& 0.
\label{bfbg_4}
\ea
\es
Furthermore,
$\Gamma\left(\gamma_5\right)$ is of the form of 
the function analyzed in Appendix~\ref{app:math},
with
\begin{equation}
	k = \frac{1}{4},\quad
	p = 9,\quad
	s = 20,\quad
	w = 2\lambda_5,\quad
	v = \lambda_2,\quad
	\alpha = 0,\quad
	\beta = \frac{9}{20}.
\end{equation}
Therefore,
for it to be non-negative everywhere inside the $\gamma_5$ domain
$\left[ 0, 9/20 \right]$ it is necessary to require
\bs
\ba
\lambda_3
- \frac{3}{4}\left| \lambda_4 \right| + \sqrt{\lambda_1 \lambda_2} &\geq& 0,
\\
\lambda_3 + \sqrt{\lambda_1 
\left(\lambda_2 + \frac{9 \lambda_5}{10}\right)} &\geq& 0.
\ea
\es
Furthermore,
one must \emph{exclude} the situation where
\bs
\ba
\lambda_5 &<& -\frac{5 \lambda_4^2}{8 \lambda_1},
\\
\lambda_5 &<& - \frac{5}{6}\,
\sqrt{\frac{\lambda_2}{\lambda_1}} \left| \lambda_4 \right|,
\\
\lambda_3 &<&
- \sqrt{\frac{\left(9\lambda_5 + 10\lambda_2 \right)
\left(8\lambda_1 \lambda_5 + 5 \lambda_4^2 \right)}{80\lambda_5}}.
\ea
\es

\subsection{The case $n=5$}

For $n=5$ the functions that we need to minimize
depend on both $\gamma_5$ and $\gamma_6$ and read
\bs
\ba
\Xi\left(\gamma_5,\gamma_6\right) &=&
\lambda_2 + 2\lambda_5 \gamma_5 + 2\lambda_6 \gamma_6,\\
\Gamma\left(\gamma_5,\gamma_6\right) &=&
\lambda_3 - \frac{\sqrt{4 - 7\gamma_5 - 10\gamma_6}}{2} \left|\lambda_4\right| 
+ \sqrt{ \lambda_1}\,
\sqrt{\lambda_2 + 2\lambda_5 \gamma_5 + 2\lambda_6 \gamma_6}.
\label{63b}
\ea
\es
In this case the phase space is two-dimensional. 
Its boundary consists of three straight-line segments.
From the monotonicity of $\Xi\left(\gamma_5,\gamma_6\right)$,
it follows that it is sufficient to ensure that it is non-negative 
at the three vertices specified in Eq.~\eqref{vertices}.
Thus,
\bs
\label{bfbg_5}
\ba
\lambda_2 &\geq& 0,
\\
\lambda_2 + \frac{8}{7}\, \lambda_5 &\geq& 0,
\\
\lambda_2 + \frac{2}{35}\, \lambda_5^{\prime} &\geq& 0,
\ea
\es
where we conveniently introduced
\begin{equation}
	\lambda_5^{\prime} \equiv 10\lambda_5 + 7\lambda_6.
\end{equation}

Regarding $\Gamma\left(\gamma_5,\gamma_6\right)$ in Eq.~\eqref{63b},
we begin by observing that 
that function does not admit any extremum in the interior of the phase space,
since the system of equations
\be
\frac{\partial \,\Gamma}{\partial \gamma_5}
= \frac{\partial \,\Gamma}{\partial \gamma_6}
= 0
\ee
has no solution. Therefore, if there is a minimum, it 
must reside at one of the 
boundaries of the phase space
defined by Eqs.~\eqref{side1},
\eqref{side2},
and~\eqref{side3}.

At the boundary~\eqref{side1}, we set $\gamma_6 = 0$ and
$\Gamma\left(\gamma_5,\gamma_6\right)$ becomes
\ba
\Gamma\left(\gamma_5,0\right) &=&
\lambda_3 - \frac{\sqrt{4 - 7\gamma_5}}{2} \left|\lambda_4\right| 
+\sqrt{ \lambda_1}\, \sqrt{\lambda_2 + 2\lambda_5 \gamma_5}.
\label{n51}
\ea
Equation~\eqref{n51} has the form of the function studied
in Appendix~\ref{app:math} with
\begin{equation}
	k = \frac{1}{2},\quad
	p = 4,\quad
	s = 7,\quad
	w = 2\lambda_5,\quad
	v = \lambda_2,\quad
	\alpha = 0,\quad
	\beta = \frac{4}{7}.
\end{equation}
Therefore,
$\Gamma\left(\gamma_5,0\right)$ is non-negative
for all $\gamma_5 \in \left[ 0 , 4/7 \right]$ if we assume
\bs
\label{bgbf_51}
\ba
\lambda_3 - \left| \lambda_4 \right|
+ \sqrt{\lambda_1 \lambda_2} &\geq& 0,
\\
\lambda_3 + \sqrt{\lambda_1 
\left(\lambda_2 + \frac{8}{7}\lambda_5\right)} &\geq& 0,
\label{69b}
\ea
\es
and moreover if we exclude any situation where
\bs
\label{70}
\ba
\lambda_5 &<& -\frac{7 \lambda_4^2}{8 \lambda_1}
\\
\lambda_5 &<&
-\frac{7}{8}\, \sqrt{\frac{\lambda_2}{\lambda_1}}
\left| \lambda_4 \right|,
\\
\lambda_3 &<&
- \sqrt{\frac{\left(8\lambda_5 + 7\lambda_2 \right)
    \left(8\lambda_5 \lambda_1 + 7 \lambda_4^2 \right)}{56\lambda_5}}.
\ea
\es
%

At the boundary~\eqref{side2} we take $\gamma_5 = \left( 10/7 \right) \gamma_6$
to obtain
\ba
\Gamma\left(\frac{10}{7}\gamma_6, \gamma_6\right) &=&
\lambda_3 - \left|\lambda_4\right| \sqrt{1 - 5\gamma_6}
+\sqrt{ \lambda_1}\, \sqrt{\lambda_2 + \frac{2}{7}\, \lambda_5^{\prime} \gamma_6}.
\label{n52}
\ea
This function has the form of the function studied in Appendix~\ref{app:math},
with
\begin{equation}
	k = 1,\quad
	p = 1,\quad
	s = 5,\quad
	w = \frac{2}{7}\, \lambda_5^{\prime},\quad
	v = \lambda_2,\quad
	\alpha = 0,\quad
	\beta = \frac{1}{5},
\end{equation}
Therefore,
it is non-negative $\forall \gamma_6 \in \left[ 0 , 1/5 \right]$ if we require
\bs
\label{bgbf_52}
\ba
\lambda_3 - \left| \lambda_4 \right| + \sqrt{\lambda_1 \lambda_2} &\geq& 0,
\\
\lambda_3 + \sqrt{\lambda_1 
\left(\lambda_2 + \frac{2}{35}\, \lambda_5^{\prime}\right)} &\geq& 0,
\label{73b}
\ea
\es
and if we exclude the situation where
\bs
\label{74}
\ba
\lambda_5^{\prime} &<& -\frac{35 \lambda_4^2}{2 \lambda_1},
\\
\lambda_5^{\prime} &<&
-\frac{35}{2}\sqrt{\frac{\lambda_2}{\lambda_1}} \left| \lambda_4 \right|,
\\
\lambda_3 &<&
- \sqrt{\frac{\left(2\lambda_5^{\prime} + 35\lambda_2 \right)
\left(2\lambda_5^{\prime} \lambda_1 + 35 \lambda_4^2 \right)}
{70\lambda_5^{\prime}}}.
\ea
\es
%

At the boundary~\eqref{side3}
the function $\Gamma\left(\gamma_5,\gamma_6\right)$ takes the form
\be
\Gamma\left(-\frac{10}{7}\gamma_6 + \frac{4}{7},\, \gamma_6\right) =
\lambda_3 +
\sqrt{\lambda_1}\,
\sqrt{\lambda_2 + \frac{8}{7}\lambda_5 + 
 \left( -\frac{20}{7}\lambda_5 + 2 \lambda_6 \right) \gamma_6}.
\label{n53}
\ee
This function is monotonic in $\gamma_6$
and the n\&s conditions for it to be non-negative
$\forall \gamma_6 \in \left[ 0 , 1/5 \right]$
are conditions~\eqref{69b} and~\eqref{73b}.

To summarize,
the n\&s BFB conditions for the case $n=5$ are~\eqref{bfbg_5},
\eqref{bgbf_51},
\eqref{bgbf_52},
and besides one must impede both conditions~\eqref{70}
and conditions~\eqref{74}.

\subsection{The case $n=6$}
\label{sec:BFBn6}

When $n=6$,
the functions that we need to minimize read
\bs
\ba
\Xi\left(\gamma_5,\gamma_6\right) &=&
\lambda_2 + 2\lambda_5 \gamma_5 + 2\lambda_6 \gamma_6,\\
\Gamma\left(\gamma_5,\gamma_6\right) &=&
\lambda_3
- \frac{\sqrt{25 - 36\gamma_5 - 56\gamma_6}}{4} \left|\lambda_4\right|
+\sqrt{ \lambda_1}\,
\sqrt{\lambda_2 + 2\lambda_5 \gamma_5 + 2\lambda_6 \gamma_6}.
\ea
\es
Neither of these functions admits extrema inside the phase space,
so their minima must reside at its boundary.
That boundary has two curved segments,
\textit{viz.} Eqs.~\eqref{3nvjfod} and~\eqref{parametric}.
We firstly require the two functions
to be non-negative at the vertices of the phase space,
given in Eqs.~\eqref{vert_6}.
The following \emph{necessary} BFB conditions are then found:
\bs
\label{necessary_g}
\ba
\lambda_2 &\geq& 0,\\
\lambda_2^{(1)} \equiv \lambda_2 + \frac{8}{9}\, \lambda_5 &\geq& 0,
\label{xi6v1}\\
\lambda_2^{(2)} \equiv \lambda_2 + 2 \varsigma \lambda_5
+ 2 \varrho \lambda_6 &\geq& 0,
\label{xi6v2}\\
\lambda_2^{(3)} \equiv \lambda_2 + \frac{89}{90}\, \lambda_5
+ \frac{9}{35}\, \lambda_6 &\geq& 0,\\
\lambda_2^{(4)} \equiv \lambda_2 + \frac{5}{18}\, \lambda_5
+ \frac{5}{7}\, \lambda_6 &\geq& 0,
\ea
\es
and
\bs
\label{necessary_f}
\ba
\lambda_3 + \sqrt{\lambda_1\lambda_2} - \frac{5}{4} \left| \lambda_4 \right| 
&\geq&  0,\\
\lambda_3 + \sqrt{\lambda_1 \lambda_2^{(1)}} 
- \frac{3}{4} \left| \lambda_4 \right|
&\geq&  0, \label{gama6v1}\\
\lambda_3 + \sqrt{\lambda_1 \lambda_2^{(2)}} 
- \frac{5\left( 24\sqrt{5} -41 \right)}{436} \left| \lambda_4 \right| 
&\geq&  0, \label{gama6v2}\\
\lambda_3 + \sqrt{\lambda_1 \lambda_2^{(3)}}
&\geq&  0, \label{q3}
\\
\lambda_3 + \sqrt{\lambda_1 \lambda_2^{(4)}} 
&\geq&  0. \label{q4}
\ea
\es

If we ignore the slight concavity of segment~\eqref{parametric}
and instead describe the boundary connecting vertices $V_2$ and $V_3$
by the linear Eq.~\eqref{parametric_approx},\footnote{This 
implies that we are exploring a phase space slightly larger than 
the true one, and as a result, the BFB conditions we impose are, 
strictly speaking, necessary instead of n\&s conditions. 
Nevertheless, we have 
verified that this approximation has a negligible practical impact. We 
generated approximately $3.6 \times 10^9$ random sets of $\lambda_i$ 
within the ranges $\lambda_{1,2} \in [0, 4\pi]$ and 
$\lambda_{3,4,5,6} \in [-4\pi, 4\pi]$. Among these, around 
$10^9$ sets satisfied our necessary BFB conditions. Notably, 
we did not encounter a single case where using the true 
concave boundary, instead of its straight approximation, would 
have changed our conclusions regarding any of these $10^9$ points.}
then the boundary of phase space consists of four straight segments
and just one curved one.
The function $\Xi\left(\gamma_5,\gamma_6\right)$ is monotonic
along a straight segment,
therefore the four straight segments do not generate any further
BFB condition besides conditions~\eqref{necessary_g}.
At the curved segment~\eqref{3nvjfod} one has
\ba
\label{8fdfuf9}
\widetilde \Xi \left( \lambda_6 \right) \equiv
\Xi \left( \frac{2 \left( 1 + 7 \gamma_6 + \sqrt{1 - 7 \gamma_6} \right)}{9},\,
\gamma_6 \right) &=&
\left( \lambda_2 + \frac{4}{9}\, \lambda_5 \right)
+ \left( \frac{28}{9}\, \lambda_5 + 2 \lambda_6 \right) \gamma_6
\no & &
+ \frac{4 \lambda_5}{9}\, \sqrt{1 - 7 \gamma_6},
\ea
which must be non-negative $\forall \gamma_6 \in \left[ 0, \varrho \right]$.
One must avoid the situation where
$\widetilde \Xi \left( \lambda_6 \right)$
has a minimum $\mu \in \left[0, \varrho \right]$
with $\widetilde \Xi \left( \mu \right) < 0$.
This means that one must \emph{impede} the situation where
\bs
\label{xi6vinfty}
\ba
7 \lambda_5 + 9 \lambda_6 &<& 0,
\\
7 \left( 5\, \sqrt{5} - 8 \right) \lambda_5 - 54 \lambda_6 &<& 0,
\\
\lambda_2^{(1)} + 
\frac{2 \left( 7 \lambda_5 + 9 \lambda_6 \right)^2}{63 \Lambda}&<& 0,
\label{eq:eeeewew}
\ea
\es
where 
\be 
\Lambda \equiv 14 \lambda_5 + 9 \lambda_6. 
\ee

Since $\gamma_5$ and $\gamma_6$ are linearly dependent along a straight segment,
$\Gamma(\gamma_5, \gamma_6)$ can be cast
into the form of the function analyzed in Appendix~\ref{app:math}
for each one of the four straight segments of the boundary.
The parameters are
\bs
\label{paramsss}
\ba
\overline{V_0\,V_1}:&&
	k = \frac{1}{4},\quad
	p = 25,\quad
	s = 36,\quad
	w = 2 \lambda_5,\quad
	v = \lambda_2,\quad
	\alpha = 0,\quad
	\beta = \frac{4}{9},
        \label{paramsss1} \\
\overline{V_2\,V_3}:&&
	k = \frac{1}{4},\quad
	p = \frac{225 \left( 2\,999 - 912\, \sqrt{5} \right)}{39\,961},\quad
	s = \frac{70}{9}\, p,\no          
	& & w = \frac{7 \left( 71\,677 - 38\,000\,
          \sqrt{5} \right)}{119\,883}\, \lambda_5
        + 2 \lambda_6,\quad
	v = \lambda_2 + \frac{25 \left( 6\,485 + 4\,104\, \sqrt{5}
          \right)}{359\,649}\, \lambda_5,\no
	&&\alpha = \varrho,\quad
	\beta = \frac{9}{70},
        \label{paramsss2} \\
\overline{V_3\,V_4}:&&
	k = 0,\quad
	w = - \frac{28}{9}\, \lambda_5 + 2 \lambda_6,\quad
	v = \lambda_2 + \frac{25}{18}\, \lambda_5,\quad
	\alpha = \frac{9}{70},\quad
	\beta = \frac{5}{14},\\
\overline{V_0\,V_4}:&&
	k = \frac{1}{4},\quad
	p = 25,\quad
	s = 70,\quad
	w = \frac{7}{9} \lambda_5 + 2\lambda_6,\quad
	v = \lambda_2,\quad
	\alpha = 0,\quad
	\beta = \frac{5}{14}.
\hspace*{9mm} \label{paramsss4}
\ea
\es
The straight segment connecting $V_3$ to $V_4$ has $k=0$,
so conditions~\eqref{q3} and~\eqref{q4}
are n\&s to guarantee that $\Gamma(\gamma_5, \gamma_6) \ge 0$
everywhere on that segment.
For the other three cases,
one should exclude any values of
$\lambda_1, \ldots, \lambda_6$ 
where all the conditions~\eqref{B9},
\eqref{B10},
and~\eqref{B11} are satisfied---for each set of parameters 
$\left\{ k, p, s, v, w, \alpha, \beta\right\}$ in Eqs.~\eqref{paramsss1},
\eqref{paramsss2},
and~\eqref{paramsss4}.
Notice that $p - s \beta = 0$
for the segments $\overline{V_2\,V_3}$ and $\overline{V_0\,V_4}$,
therefore for those segments condition~\eqref{B10b} automatically holds.

On the convex segment connecting $V_1$ and $V_2$ we introduce
\be
x \equiv \sqrt{1-7\gamma_6}, \label{x}
\ee
with $x \in \left[ \sqrt{1-7\varrho}, 1 \right]$.
Then, the functions that we ought to minimize on this 
segment are\footnote{Minimizing $g \left( x \right)$
and $f \left( x \right)$ with respect to $x$
is equivalent to minimizing the functions in the left-hand side
of Eqs.~\eqref{vmf9} with respect to $\gamma_6$
because the transformation $\gamma_6 \to x$ in Eq.~\eqref{x}
is \emph{injective} in the domain $\gamma_6 \in \left[ 0, \varrho \right]
\Leftrightarrow	
x \in \left[ \sqrt{1-7\varrho}, 1 \right]$.}
\bs
\label{vmf9}
\ba
\Xi \left(
\frac{2 \left( 1 + 7 \gamma_6 + \sqrt{1 - 7 \gamma_6} \right)}{9},\,
\gamma_6 \right) &=&
g(x) \equiv
\left( \lambda_2 + \frac{4 \lambda_5}{9} + \frac{2 \Lambda }{63}\right)
+\frac{4 \lambda_5}{9}\, x 
-\frac{2\Lambda}{63}\,  x^2, \hspace*{5mm}
\\
\Gamma \left(
\frac{2 \left( 1 + 7 \gamma_6 + \sqrt{1 - 7 \gamma_6} \right)}{9},\,
\gamma_6 \right) &=&
f(x) \equiv
\lambda_3 - \frac{4x-1}{4} \left| \lambda_4 \right|
+\sqrt{\lambda_1 \, g(x)}.
\ea
\es
Their first and second derivatives read
\bs
\ba
g^\prime(x) &=&
\frac{4 \lambda_5}{9} -\frac{4\Lambda }{63}\, x,\\
g^{\prime\prime}(x) &=& 
-\frac{4\Lambda }{63},\\
f^\prime(x) &=&
-|\lambda_4|
+\frac{g^\prime(x)}{2}
\sqrt{\frac{\lambda_1}{g(x)}},\\
f^{\prime\prime}(x) &=& 
\frac{2\, g(x)\, g^{\prime\prime}(x)-  \left[ g^\prime(x) \right]^2}{4}\,
\sqrt{\frac{\lambda_1}{g(x)^3}}.
\ea
\es

We have already seen that for $g(x)$ to be non-negative for all $x$
in the interval $\left[ \sqrt{1 - 7 \varrho},\, 1 \right]$
one just needs to enforce conditions~\eqref{xi6v1}
and~\eqref{xi6v2} while avoiding conditions~\eqref{xi6vinfty}.
For the function $f(x)$ to be non-negative everywhere in the same interval
one must enforce conditions~\eqref{gama6v1} and~\eqref{gama6v2}
while avoiding the situation where
\be
f^\prime (\sqrt{1-7\varrho}) < 0, \quad
f^\prime (1) > 0, \quad
\mbox{and} \quad \exists \mu: \quad
f^\prime (\mu) = 0, \quad
f(\mu) < 0.
\label{cond_f_0}
\ee
The third condition~\eqref{cond_f_0} yields the solution
\be
\mu = \frac{ 7\lambda_5}{\Lambda}
- \frac{9 \left| \lambda_4 \right|}{\Lambda}\, \sqrt{\frac{7K}{2S}},
\label{deriv_f_mu}
\ee
which only exists if $K / S > 0$,
where
\bs
\ba
K &\equiv& 2 \left( 7\lambda_5 + 3\lambda_6 \right)^2 + 7\lambda_2\Lambda,
\\
S &\equiv& 63 \lambda_4^2 + 2 \lambda_1 \Lambda.
\ea
\es
When the first and second conditions~\eqref{cond_f_0} are satisfied,
$\mu$ is guaranteed to be a minimum
and to lie inside the domain $\left[ \sqrt{1-7\varrho},\, 1 \right]$.
Under these assumptions, the positivity of $f^{\prime \prime} ( \mu )$
further specifies that
\bs
\ba
K &<& 0,
\\
S &<& 0.
\ea
\es
The remaining three conditions~\eqref{cond_f_0} yield, respectively
\bs
\ba
- \left| \lambda_4 \right|
+ \frac{2 \left( 7 \lambda_5 - \Lambda\, \sqrt{1-7\varrho} \right)}{63}\,
\sqrt{\frac{\lambda_1}{\lambda_2^{(2)}}} &<& 0,
\\
- \left| \lambda_4 \right| + \frac{2 \left( 7\lambda_5 - \Lambda \right)}{63}\,
\sqrt{\frac{\lambda_1}{\lambda_2^{(1)}}} &>& 0,
\\
\lambda_3 - \frac{4 \mu-1}{4} \left| \lambda_4 \right|
+\sqrt{\lambda_1 \, g(\mu)} &<& 0.
\ea
\es
Therefore, besides enforcing conditions~\eqref{gama6v1} and \eqref{gama6v2},
one must avoid any situation where all the following conditions hold:
\bs
\label{exlude_convex_6}
\ba
\Lambda &<& -\frac{2 \left( 7\lambda_5 + 3\lambda_6 \right)^2}{7 \lambda_2},\\
\Lambda &<& -\frac{63 \lambda_4^2}{2 \lambda_1},\\
\left| \lambda_4 \right| &>&  \frac{2}{63}\,
\sqrt{\frac{\lambda_1}{\lambda_2^{(2)}}}\,
\left(  7\lambda_5 - \Lambda\, \sqrt{1-7\varrho}\right),
\label{2mdvkfo} \\
\left| \lambda_4 \right| &<& - \frac{2}{63}\,
\sqrt{\frac{\lambda_1}{\lambda_2^{(1)}}}\,
\left( 7\lambda_5 + 9 \lambda_6 \right),\\
\lambda_3 
&<& \frac{4 \mu-1}{4} \left| \lambda_4 \right| -\sqrt{\lambda_1 \, g(\mu)}.
\ea
\es
In condition~\eqref{2mdvkfo},
note that $\sqrt{1 - 7 \varrho}
= \left. 6\left(5\, \sqrt{5} - 4 \right) \right/ 109$.

To summarize, the necessary BFB conditions for the case 
$n=6$ are~\eqref{necessary_g} and \eqref{necessary_f}.
Besides that, it is sufficient to \textit{exclude} any situation where
Eqs.~\eqref{xi6vinfty} or Eqs.~\eqref{exlude_convex_6} hold,
or where conditions~\eqref{B9}, \eqref{B10},
and~\eqref{B11} are satisfied for each set of parameters 
$\left\{ k, p, s, v, w, \alpha, \beta\right\}$ in Eqs.~\eqref{paramsss1},
\eqref{paramsss2},
and~\eqref{paramsss4}.

\section{Vacuum stability}
\label{sec:vacuumstability}

Following Ref.~\cite{Milagre:2024nvy},
we classify the possible extrema of the scalar potential (SP)
of our model in the following exhaustive way:
\begin{itemize}
\item The type-0 extremum has
  \be
  \label{3}
  \left\langle \Phi \right\rangle = 0,
  \quad
  \left\langle \Delta_n \right\rangle = 0.
  \ee
\item A type-I extremum has
  \be
  \label{4}
  \left\langle \Phi \right\rangle \neq 0,
  \quad
  \left\langle \Delta_n \right\rangle = 0.
  \ee
\item A type-II extremum has
  \be
  \label{5}
  \left\langle \Phi \right\rangle = 0,
  \quad
  \left\langle \Delta_n \right\rangle \neq 0.
  \ee
\item A type-III extremum has
  \be
  \label{6}
  \left\langle \Phi \right\rangle \neq 0,
  \quad
  \left\langle \Delta_n \right\rangle \neq 0.
  \ee
\end{itemize}
In Ref.~\cite{Milagre:2024nvy} two of us have proved that,
if the SP of a renormalizable $SU(2)$-symmetric model
with a scalar doublet $\Phi$ contains neither linear terms,
nor trilinear terms,
nor quadratic terms with two different multiplets,
and if moreover $\Phi$ only interacts
with any other scalar $SU(2)$ multiplet $\Delta_n$
through quartic invariants constructed out of
\be
\Phi \otimes \widetilde{\Phi} \otimes
\Delta_n \otimes \widetilde{\Delta}_n
\label{pppp}
\ee
---where $\widetilde{\Delta}_n$ stands for the $SU(2)$ multiplet
with the same dimension of $\Delta_n$ and formed by the complex conjugates
of the scalar fields of $\Delta_n$---then a type-I
\emph{local minimum} of the SP
has a lower value of the SP than any type-0 or type-III extremum.
As a consequence,
either the \emph{global minimum} of the SP is that type-I local minimum,
or it is a type-II minimum (if there is any).
Since the models studied in this work fulfil the assumptions of that theorem,
in order to assure the stability of the type-I vacuum
one just has to find the conditions which ensure that:
\begin{enumerate}
\item The vacuum is a local minimum of the SP.
  This is equivalent to
  all the physical scalars having positive masses-squared.
\item No type-II extremum has an expectation value of the SP lower than
  the one of the vacuum.
\end{enumerate}
We concentrate here on the second task.

\subsection{Type-I vacuum}
The expectation value of the SP at the type-I vacuum
is independent of the dimension of $\Delta_n$;
it is given by
\be
\overline{V}_{\mathrm{I}} = \mu_1^2 \left\langle F_1\right\rangle  
+ \frac{\lambda_1}{2} \left\langle F_1 \right\rangle ^2,
\label{ivufi4}
\ee
where $\left\langle F_1 \right\rangle$ is the expectation value
of the $SU(2)$ invariant defined in Eq.~\eqref{F1}.
The stationarity condition for $\overline{V}_{\mathrm{I}}$ is
\be
\mu_1^2 = - \lambda_{1} \left\langle F_1 \right\rangle.
\label{39idif}
\ee
Boundedness from below of the SP requires $\lambda_1$ to be positive.
Hence,
Eq.~\eqref{39idif} tells us that $\mu_1^2$ must be negative
for the type-I vacuum to exist.
Equations~\eqref{ivufi4} and~\eqref{39idif} imply that
\be
\overline V_\mathrm{I}
= - \frac{\left( \mu_1^2 \right)^2}{2 \lambda_{1}}.
\ee

\subsection{Type-II extremum}

At a type-II extremum, only $\Delta_n$ acquires a VEV.
According to Eq.~\eqref{THESP} with $F_1 = 0$, 
the expectation value of the SP is
\be
\overline{V}_{\mathrm{II}} = 
\mu_2^2 \left\langle F_2\right\rangle  
+ \frac{\left\langle\Xi\left(\gamma_5, \gamma_6 \right)\right\rangle}{2}
\left\langle F_2 \right\rangle^2 ,
\label{POTII}
\ee
where $\left\langle F_2 \right\rangle$ and 
$\left\langle\Xi\left(\gamma_5, \gamma_6 \right)\right\rangle$
are the expectation values of the $SU(2)$ invariants
$F_2$ and $\Xi\left(\gamma_5, \gamma_6 \right)$ 
at that type-II extremum,
respectively.
Further note that
\be
\left\langle\Xi\left(\gamma_5, \gamma_6 \right)\right\rangle =
\left\{ \begin{array}{l}
  \displaystyle{\lambda_2 \quad \Leftarrow
    \quad \mbox{either}\ n=1\ \mbox{or}\ n=2},
  \\
  \displaystyle{\lambda_2 + 2\lambda_5 \left\langle \gamma_5 \right\rangle
    \quad \Leftarrow
    \quad \mbox{either}\ n=3\ \mbox{or}\ n=4,}
  \\
  \displaystyle{\lambda_2 + 2\lambda_5 \left\langle \gamma_5 \right\rangle
    + 2\lambda_6 \left\langle \gamma_6 \right\rangle
    \quad \Leftarrow \quad \mbox{either}\  n=5\ \mbox{or}\ n=6.}
\end{array} \right.
\ee
Effecting the minimization of $\overline{V}_{\mathrm{II}}$ in Eq.~\eqref{POTII}
relative to $\left\langle F_2 \right\rangle$,
one finds that at a type-II extremum
\be
\mu_2^2 = -
\left\langle\Xi \left( \gamma_5, \gamma_6 \right)\right\rangle
\left\langle F_2 \right\rangle.
\label{POTII2}
\ee
The SP being bounded from below necessitates that 
$\Xi\left(\gamma_5, \gamma_6 \right)$ is non-negative,
as we have extensively discussed in the previous section.
Therefore,
a type-II extremum only exists if $\mu_2^2 < 0$.
From Eqs.~\eqref{POTII} and~\eqref{POTII2},
\be
\overline V_\mathrm{II}
= - \frac{\left( \mu_2^2 \right)^2}{2
  \left\langle\Xi \left( \gamma_5, \gamma_6 \right)\right\rangle}.
\ee
To ensure that the type-I vacuum lies below the type-II extremum
with the lowest $\overline V_\mathrm{II}$,
we require the parameters of the SP to be such that
all possible
$\left\langle\Xi \left( \gamma_5, \gamma_6 \right)\right\rangle$ satisfy
\be
\left\langle\Xi \left( \gamma_5, \gamma_6 \right)\right\rangle 
> \lambda_1 \left( \frac{\mu_2^2}{\mu_1^2} \right)^2.
\label{condition}
\ee

\subsection{Conditions for vacuum stability}

When $\Delta_n$ is either a singlet ($n=1$) or a doublet ($n=2$),
$\left\langle\Xi \left( \gamma_5, \gamma_6 \right)\right\rangle = \lambda_2$ 
and condition~\eqref{condition} simply implies that
\be
\label{simple2}
\lambda_2 >  \left( \frac{\mu_2^2}{\mu_1^2} \right)^2 \lambda_1.
\ee
This condition for the type-I extremum to be the global minimum
of the SP is consistent with the results derived for the 
complex singlet extension of the SM in \cite{Barger:2008jx}, 
and with the analogous condition found in 
\cite{Branco:2011iw, Ginzburg:2010wa, Dercks:2018wch} for the 
$U(1)$-symmetric 2HDM.

For larger $n$,
$\left\langle\Xi \left( \gamma_5, \gamma_6 \right)\right\rangle$ 
may take several values
and we must investigate the space spanned by
$\gamma_5$ and $\gamma_6$.
The task of finding the values of 
$\left\langle\Xi \left( \gamma_5, \gamma_6 \right)\right\rangle$
that lead to the smallest possible value of $\overline{V}_{\mathrm{II}}$
can be simplified by noting that
$\left\langle\Xi \left( \gamma_5, \gamma_6 \right)\right\rangle$
is linear in $\left\langle\gamma_5\right\rangle$
and $\left\langle\gamma_6\right\rangle$.
Therefore,
any type-II extremum must lie at the \emph{boundary}
of the phase space~\cite{degee}.
Furthermore,
the minimum of the potential is attained at 
the points of phase space that extend the farthest in some direction.
Hence,
when the boundary consists of straight-line segments meeting at vertices,
it suffices to evaluate
$\left\langle\Xi\left( \gamma_5, \gamma_6 \right)\right\rangle$
at those vertices to find the possible type-II extrema.
Consequently, 
\be
\left\langle\Xi\left( \gamma_5, \gamma_6 \right)\right\rangle 
= \left\{ \begin{array}{l}
  \displaystyle{\mathrm{either}\ \lambda_2 \
    \mathrm{or}\ \lambda_2 + \frac{2\lambda_5}{3}
    \ \Leftarrow\ n=3,}
  \\*[3mm]
  \displaystyle{\mathrm{either}\ \lambda_2 \
    \mathrm{or}\ \lambda_2 + \frac{9\lambda_5}{10}
    \ \Leftarrow\ n=4,}
  \\*[3mm]
  \displaystyle{\mathrm{either}\ \lambda_2, \
    \mathrm{or} \ \lambda_2 + \frac{8\lambda_5}{7},
    \ \mathrm{or} \ \lambda_2 + \frac{4\lambda_5}{7}+ \frac{2\lambda_6}{5}
    \ \Leftarrow\ n=5.}
\end{array} \right.
\ee
Therefore,
condition~\eqref{simple2} must hold for all $n \le 5$ and besides
one must require that
\bs
\ba
n=3 &\Rightarrow& \lambda_2
+ \frac{2 \lambda_5}{3} >  \left( \frac{\mu_2^2}{\mu_1^2} \right)^2 \lambda_1,
\label{stablen3}\\
n=4 &\Rightarrow& \lambda_2
+ \frac{9 \lambda_5}{10} >  \left( \frac{\mu_2^2}{\mu_1^2} \right)^2 \lambda_1,
\\
n=5 &\Rightarrow& \lambda_2
+ \mathrm{min} \left( \frac{8 \lambda_5}{7},\ \frac{4 \lambda_5}{7}
+ \frac{2 \lambda_6}{5} \right)
>  \left( \frac{\mu_2^2}{\mu_1^2} \right)^2 \lambda_1.
\ea
\es

An important consistency check may be performed in the case
$n=3$, where, using the notation of Eq.~\eqref{13},
the conditions given in Eqs.~\eqref{simple2} and \eqref{stablen3} 
exactly coincide with those presented in Eq.~(5.9) of
Ref.~\cite{Ferreira:2019hfk}.

For $n=6$,
determining the possible type-II extrema 
is more challenging because the boundary of the phase space
contains both straight-line segments and a convex segment.\footnote{For 
the moment, we are approximating the slightly concave parametric curve 
between the vertices $V_2$ and $V_3$ by a straight line, 
just as we did in the previous section.}
Note that $\left\langle\Xi\left( \gamma_5, \gamma_6 \right)\right\rangle$ 
is still linear in both $\left\langle \gamma_5 \right\rangle$
and $\left\langle \gamma_6 \right\rangle$,
so we introduce the direction of steepest descent
$\vec{n} = -\left( \left\langle \gamma_6 \right\rangle,\,
\left\langle \gamma_5 \right\rangle \right)$.
If $\vec{n}$ points to one of the straight boundaries,
then the farthest protruding points in the direction of $\vec{n}$
coincide with the vertices of the phase space given in Eq.~\eqref{vert_6}.
The possible type-II extrema then have
\be
\left\langle\Xi\left( \gamma_5, \gamma_6 \right)\right\rangle
= \mathrm{either}\ \lambda_2,\
\mathrm{or}\ \lambda_2 + \frac{8\lambda_5}{9},\
\mathrm{or}\ \lambda_2 + 2\varsigma\lambda_5 + 2\varrho\lambda_6,\
\mathrm{or}\ \lambda_2 + \frac{89\lambda_5}{90}+ \frac{9\lambda_6}{35},\
\mathrm{or}\ \lambda_2 + \frac{5\lambda_5}{18}+ \frac{5\lambda_6}{7}.
\ee
However,
if $\vec{n}$ points to the boundary connecting the vertices $V_1$ and $V_2$, 
then the farthest protruding point in the direction of $\vec{n}$
will lie somewhere on the convex boundary~\eqref{3nvjfod}.
In other words, we need to find the minimum of the function
$\widetilde \Sigma \left( \gamma_6 \right)$
defined in Eq.~\eqref{8fdfuf9}
in the range $\gamma_6 \in \left[0, \varrho \right]$.
That was done already when studying the boundedness-from-below
conditions in Section~\ref{sec:BFBn6};
more specifically, 
if
\be
7 \left( 5\, \sqrt{5} - 8 \right) \lambda_5
< 54 \lambda_6 < - 42 \lambda_5,
\ee
then, from Eqs.~\eqref{eq:eeeewew}, one more possible value of
$\left\langle\Xi\left( \gamma_5, \gamma_6 \right)\right\rangle$ is
\be
\lambda_2 + \frac{2 \left( 7 \lambda_5 + 3 \lambda_6 \right)^2}{7
  \left( 14 \lambda_5 + 9 \lambda_6 \right)}.
\ee

If we want to be more rigorous,
then we must not neglect the curvature
of the parametric curve~\eqref{parametric}.
We must then consider,
for each set of values of $\lambda_2$,
$\lambda_5$,
and $\lambda_6$ the function of $t$
\bs
\ba
\left\langle\Xi\left( \gamma_5, \gamma_6 \right)\right\rangle
\left( t \right) &=& \lambda_2 + \frac{40 t}{63 \left( 75 t^4 - 100 t^3 + 90 t^2
  + 12 t + 3 \right)}  \times
\\ & &
\bigg[ 
144 t^2 \left( - 165 t^4 + 212 t^3
  - 102 t^2 + 36 t + 3 \right)\lambda_6\\
  &&
+7 \left( 1125 t^7 - 195 t^6 + 821 t^5
   - 1371 t^4 + 823 t^3 \right.\\
   &&
   \left.+ 399 t^2 + 175 t + 15 \right) \lambda_5 \bigg],
\ea
\es
and we must look for minima of this function in the interval
\be
\frac{3}{5} \le t \le \frac{3 + 2 \sqrt{5}}{11}.
\ee
Any minima must be treated as extra possibilities for 
$\left\langle\Xi\left( \gamma_5, \gamma_6 \right)\right\rangle$.

\section{Conclusions and discussion}
\label{sec:conclusions}

In this paper,
we have considered the extension of the electroweak Standard Model
through a single scalar multiplet $\Delta_n$ of $SU(2)$.
We have assumed that multiplet to enjoy a $U(1)$ symmetry
$\Delta_n \to \Delta_n \exp \left( i \vartheta \right)$
and to have no vacuum expectation value (VEV),
so as not to perturb,
at tree level,
the successful SM prediciton~\eqref{mwmz}
(it does perturb it at loop level, though).
We have analyzed the scalar potential $V$
of this model with only two scalar multiplets---$\Delta_n$
and the SM doublet $\Phi$---in order to find out the ranges of its various
$SU(2)$-invariants, and thus,
the conditions for it to be bounded from below
(and thus be able to produce a vacuum state),
and the conditions for our preferred vacuum state
(where only $\Phi$ has VEV)
to be the global minimum of $V$.

With just one scalar $SU(2)$ multiplet $\Delta_n$ with $n$ components,
there are $m$ quartic $SU(2)$-invariants in $V$,
where $m=1$ for either $n=1$ or $n=2$,
$m=2$ for either $n=3$ or $n=4$,
$m=3$ for either $n=5$ or $n=6$,
and so on.
The increasing number of invariants
renders their ranges increasingly complicated to calculate.
As demonstrated in Sec.~\ref{sec:OS}, 
for $n \leq 4$, 
we were able to identify the phase space boundaries through 
direct algebraic manipulation of the invariants. 
To tackle the same problem for the cases $n=5$ and $n=6$, 
we first assigned random values to the scalar fields of 
$\Delta_n$ to map out the phase space, 
and only afterwards we attempted to find 
suitable combinations of non-zero fields 
to characterize the boundaries analytically. 
It is important to clarify that, although numerical sampling 
both provided the initial insight
and confirmed the final analytical result, 
our determination of the phase space 
boundaries is entirely analytical.

We have found that,
starting with $n=6$,\footnote{We have also briefly explored the case $n=7$,
but we do not report on it in this paper
(and we are not planning to do it elsewhere).}
the phase space has some curved boundaries,
at least one of which is \emph{concave}.
As shown in Eq.~\eqref{parametric}, 
this concave boundary can only be expressed 
in parametric form as a ratio of high-degree polynomials.
For practical purposes, we therefore approximate it 
throughout our analysis by using the straight-line 
segment given in Eq.~\eqref{parametric_approx}.

By using the analytic equations for all the boundaries
of the phase space for the cases $n \le 6$,
we were able to deduce analytic n\&s bounded-from-below conditions on $V$,
and also analytic n\&s conditions for our desired vacuum state---where $\Phi$,
but not $\Delta_n$,
has non-zero VEV---to be the absolute minimum of the potential.
We assessed the accuracy of all our analytical 
results by performing numerical scans over the phase space.
Note, however, that due to the straight-line approximation 
adopted for the concave boundary in the case $n = 6$, 
the phase space that we have
worked with was slightly larger than the exact one. 
As a result, the conditions found by us in this case are, 
strictly speaking, necessary instead of n\&s conditions.
To evaluate the validity of this simplification, we scanned 
$10^9$ sets of couplings $\lambda_i$
and did not find a single instance 
where applying the exact concave boundary---rather than its 
linear approximation---would have altered our conclusions. 
This demonstrates the negligible practical impact of the approximation.

  One may discuss the situation where the theory has,
  instead of the $U(1)$ symmetry
  $\Delta_n \to \Delta_n \exp \left( i \vartheta \right)$,
  the smaller (discrete) $\mathbbm{Z}_2$ symmetry $\Delta_n \to - \Delta_n$.
  In the case where $n$ is odd,
  \textit{i.e.}\ where $J$ is integer,
  that makes no difference relative to the case with $U(1)$ symmetry.
  But,
  in the case where $n$ is even
  and if additionally $\Delta_n$ has a specific hypercharge,
  there is an additional $SU(2) \times U(1)$-invariant term
  in the quartic part of the scalar potential.
  Indeed,
  if $n$ is even then the product $\Delta_n \otimes \Delta_n$
  includes an $SU(2)$ triplet
  $\left( \Delta_n \otimes \Delta_n \right)_\mathbf{3}$.
  (The boldface sub-index
  indicates the dimension of the $SU(2)$ representation.)
  Then,
  besides the $F_4$ term,
  which is the $SU(2)$-invariant in
  $\left( \Phi \otimes \widetilde \Phi \right)_\mathbf{3} \otimes
  \left( \Delta_n \otimes \widetilde \Delta_n \right)_\mathbf{3}$,
  there are further terms in the potential,
  namely either
  \be
  \left[
    \left( \Phi \otimes \widetilde \Phi \right)_\mathbf{3} \otimes
    \left( \Delta_n \otimes \Delta_n \right)_\mathbf{3} \right]_\mathbf{1}
  \label{843938}
  \ee
  if $\Delta_n$ has null hypercharge,
  or
  \be
  \left[
    \left( \widetilde \Phi \otimes \widetilde \Phi \right)_\mathbf{3} \otimes
    \left( \Delta_n \otimes \Delta_n \right)_\mathbf{3} \right]_\mathbf{1}
  \label{843939}
  \ee
  if the hypercharge of $\Delta_n$ is the same as the one of $\Phi$.
  (Besides the terms~\eqref{843938} and~\eqref{843939} there are
  their Hermitian conjugates.)
  (In the case where $n=2$,
  \textit{i.e.}\ in the 2HDM,
  the term~\eqref{843939} is,
  in the standard notation,
  the term $\left( \phi_1^\dagger \phi_2 \right)^2$ with coefficient $\lambda_5$.)
  So,
  the case with $\mathbbm{Z}_2$ symmetry is,
  when $n$ is even and $\Delta_n$ has hypercharge either $0$ or $1/2$,
  more complicated than the case with $U(1)$ symmetry,
  because there is then an additional dimensionless  parameter---with
  denominator $F_1 F_2$ just as $\delta$
  in the last Eq.~\eqref{invariants}---and therefore
  the phase space has an extra dimension.

\vspace*{5mm}
\paragraph{Acknowledgements:}
The work of A.M.\ and L.L.\ has been supported
by the Portuguese Foundation for Science and Technology (FCT)
through projects UIDB/00777/2020, UIDP/ 00777/2020, and 
through the PRR (Recovery and Resilience Plan),
within the scope of the investment ``RE-C06-i06 - Science
Plus Capacity Building", measure ``RE-C06-i06.m02 -
Reinforcement of financing for International Partnerships
in Science, Technology and Innovation of the PRR", under
the project with reference 2024.01362.CERN.
A.M. was additionally supported by FCT with PhD Grant No. 2024.01340.BD.
L.L.\ was furthermore supported by projects CERN/FIS-PAR/0002/2021
and CERN/FIS-PAR/0019/2021.
The work of D.J. received funding from the Research Council of Lithuania
(LMTLT) under Contract No. S-CERN-24-2. Part of the computations were
performed using the infrastructure of the Lithuanian Particle Physics
Consortium, under the agreement between Vilnius University and the
LMTLT, No. VS-13.

\newpage
\begin{appendix}

\setcounter{equation}{0}
\renewcommand{\theequation}{A\arabic{equation}}

\section{The definitions of $F_2$, $F_4$, $F_5$, and $F_6$}
\label{app:previous}

In this appendix, we recover relevant definitions and results
of Ref.~\cite{our}.

\subsection{$n = 1$}

If $n=1$,
\textit{i.e.}\ if $J= 0$,
then $\Delta_1$ contains a single complex scalar fields $c$,
with $I_c = 0$:
\be
\Delta_1 = \left( \begin{array}{c} c \end{array} \right),
\quad
\widetilde \Delta_1 = \left( \begin{array}{c} c^\ast \end{array} \right).
\ee
Besides the $SU(2)$-invariant $F_1$ defined in Eq.~\eqref{F1},
one may write
\be
F_2 = \left| c \right|^2 \equiv C.
\label{F2_1}
\ee
The scalar potential is given by Eq.~\eqref{7a}.

\subsection{$n = 2$}

If $n=2$,
\textit{i.e.}\ if $J= 1/2$,
then
\be
\Delta_2 = \left( \begin{array}{c} c \\ d \end{array} \right),
\quad
\widetilde \Delta_2 = \left( \begin{array}{c} d^\ast \\
  - c^\ast \end{array} \right),
\ee
where $c$ and $d$ are complex scalar fields with $I_c = - I_d = 1/2$.
We can build two $SU(2)$-invariants apart from $F_1$.
Namely
\bs
\ba
F_2 &=& C + D,
\label{F2_2}
\\
F_4 &=& \frac{\left( A - B \right) \left( C - D \right)}{4}
+ \frac{a b^\ast c^\ast d + a^\ast b c d^\ast}{2}.
\ea
\es
The SP is given by Eq.~\eqref{7b}.
This model is identical to the $U(1)$-symmetric 2HDM.
The notation that most authors use for the quartic part
of the scalar potential of the latter is \cite{Branco:2011iw}
\be
V_4 =
\frac{\bar \lambda_1}{2} \left( \Phi^\dagger \Phi \right)^2
+ \frac{\bar \lambda_2}{2} \left( \Delta_2^\dagger \Delta_2 \right)^2
+ \bar \lambda_3 \left( \Phi^\dagger \Phi \right)
\left( \Delta_2^\dagger \Delta_2 \right)
+ \bar \lambda_4 \left( \Phi^\dagger \Delta_2 \right)
\left( \Delta_2^\dagger \Phi \right).
\ee
This is the same as our notation
\be
V_4 =
\frac{\lambda_1}{2}\, F_1^2
+ \frac{\lambda_2}{2}\, F_2^2
+ \lambda_3 F_1 F_2
+ \lambda_4 F_4,
\ee
with
\be
\label{theequivalence}
\bar \lambda_1 = \lambda_1, \quad
\bar \lambda_2 = \lambda_2, \quad
\bar \lambda_3 + \frac{\bar \lambda_4}{2} = \lambda_3, \quad
2 \bar \lambda_4 = \lambda_4.
\ee

\subsection{$n = 3$}

If $n=3$,
\textit{i.e.}\ if $J=1$,
then
\be
\Delta_3 = \left( \begin{array}{c} c \\ d \\ e \end{array} \right),
\quad
\widetilde \Delta_3 = \left( \begin{array}{c} e^\ast \\ - d^\ast \\ c^\ast
\end{array} \right),
\ee
where $c$,
$d$,
and $e$ are complex scalar fields with $I_c = - I_e = 1$ and $I_d = 0$.
There are now three $SU(2)$-invariants that we can build,
namely
\bs
\ba
F_2 &=& C + D + E,
\\
F_4 &=& \frac{\left( A - B \right) \left( C - E \right)}{2}
+ \frac{a b^\ast \left( c^\ast d + d^\ast e \right)
  + a^\ast b \left( c d^\ast + d e^\ast \right)}{\sqrt{2}},
\\
F_5 &=& \frac{\left| 2 c e - d^2 \right|^2}{3}.
\ea
\es
The SP is given by Eq.~\eqref{7c}.

The triplet $\Delta_3$ is more usually written in the form
\be
\Delta = \left( \begin{array}{cc} - d \left/ \sqrt{2} \right. & c
  \\ - e & d \left/ \sqrt{2} \right. \end{array} \right),
\ee
wherewith the potential is written as
\bs
\label{13}
\ba
V &=&
\mu_1^2 \Phi^\dagger \Phi
+ \mu_2^2\, \mathrm{tr} \left( \Delta^\dagger \Delta \right)
+ \frac{\lambda_1}{2} \left( \Phi^\dagger \Phi \right)^2
+ \frac{\overline{\lambda}_2}{2}
\left[ \mathrm{tr} \left( \Delta^\dagger \Delta \right) \right]^2
\\ & &
+ \overline{\lambda}_3\, \mathrm{tr} \left(
\Delta^\dagger \Delta \Delta^\dagger \Delta \right)
+ \overline{\lambda}_4 \left( \Phi^\dagger \Phi \right) \mathrm{tr} \left(
\Delta^\dagger \Delta \right)
+ \overline{\lambda}_5\, \Phi^\dagger \Delta \Delta^\dagger \Phi.
\ea
\es
The SP of Eq.~\eqref{13} is equivalent to the one of Eq.~\eqref{7c}
with 
\be
	\overline{\lambda}_2 + 2\overline{\lambda}_3 = \lambda_2, 
	\quad
	\overline{\lambda}_4 + \frac{1}{2}\overline{\lambda}_5 = \lambda_3,
	\quad
	\overline{\lambda}_5 = \lambda_4,
	\quad
	-\frac{2}{3}\overline{\lambda}_3 = \lambda_5.
\label{subssssn3}
\ee

\subsection{$n = 4$}

If $n=4$,
\textit{i.e.}\ if $J=3/2$,
then
\be
\Delta_4 = \left( \begin{array}{c} c \\ d \\ e \\ f \end{array} \right),
\quad
\widetilde \Delta_4 = \left( \begin{array}{c} f^\ast \\ - e^\ast \\ d^\ast \\
  - c^\ast \end{array} \right),
\ee
where $c$,
$d$,
$e$,
and $f$ are complex scalar fields with $I_c = - I_f = 3/2$
and $I_d = - I_e = 1/2$.
The SP is still the one in Eq.~\eqref{7c},
now with
\bs
\ba
F_2 &=& C + D + E + F,
\\
F_4 &=& \frac{\left( A - B \right) \left( 3 C + D - E - 3 F \right)}{4}
\\ & &
+ \frac{a b^\ast \left( \sqrt{3}\, c^\ast d + 2 d^\ast e
  + \sqrt{3}\, e^\ast f \right) + \mathrm{H.c.}}{2},
\\
F_5 &=& \frac{2 \left| \sqrt{3}\, c e - d^2 \right|^2
  + 2 \left| \sqrt{3}\, d f - e^2 \right|^2
  + \left| 3 c f - d e \right|^2}{5}.
\ea
\es

\subsection{$n = 5$}

If $n=5$,
\textit{i.e.}\ if $J=2$,
then
\be
\label{mvfp5}
\Delta_5 = \left( \begin{array}{c} c \\ d \\ e \\ f \\ g \end{array} \right),
\quad
\widetilde \Delta_5 = \left( \begin{array}{c} g^\ast \\ - f^\ast \\ e^\ast \\
  - d^\ast \\ c^\ast \end{array} \right),
\ee
where $c$,
$d$,
$e$,
$f$,
and $g$ are complex scalar fields with $I_c = - I_g = 2$,
$I_d = - I_f = 1$,
and $I_e = 0$.
The set of $SU(2)$-invariants now reads
\bs
\label{A17}
\ba
F_2 &=& C + D + E + F + G,
\\
F_4 &=& \frac{\left( A - B \right) \left( 2 C + D - F - 2 G \right)}{2}
\\ & &
+ \left\{ a b^\ast \left[ c^\ast d + f^\ast g
  + \sqrt{\frac{3}{2}} \left( d^\ast e + e^\ast f \right) \right]
ll  + \mathrm{H.c.} \right\},
\\
F_5 &=& \frac{\left| 2 \sqrt{2}\, c e - \sqrt{3}\, d^2 \right|^2
  + \left| 2 \sqrt{2}\, e g - \sqrt{3}\, f^2 \right|^2}{7}
\\ & &
  + \frac{2 \left( \left| \sqrt{6}\, c f - d e \right|^2
  + \left| \sqrt{6}\,  d g - e f \right|^2
  + \left| 2 c g + d f - e^2 \right|^2 \right)}{7},
\\
F_6 &=& \frac{\left| 2 c g - 2 d f + e^2 \right|^2}{5},
\ea
\es
and the full SP takes the form in Eq.~\eqref{7d}.

\subsection{$n = 6$}

If $n=6$,
\textit{i.e.}\ if $J=5/2$,
then
\be
\label{mvfp6}
\Delta_6 = \left( \begin{array}{c} c \\ d \\ e \\ f \\ g \\ h
\end{array} \right),
\quad
\widetilde \Delta_6 = \left( \begin{array}{c} h^\ast \\ - g^\ast \\ f^\ast \\
  - e^\ast \\ d^\ast \\ - c^\ast \end{array} \right),
\ee
where $c$,
$d$,
$e$,
$f$,
$g$,
and $h$ are complex scalar fields with $I_c = - I_h = 5/2$,
$I_d = - I_g = 3/2$,
and $I_e = - I_f = 1/2$.
The SP is the one of Eq.~\eqref{7d},
with
\bs
\label{A20}
\ba
F_2 &=& C + D + E + F + G + H,
\\
F_4 &=& \frac{\left( A - B \right)
  \left( 5 C + 3 D + E - F - 3 G - 5 H \right)}{4}
\\ & &
+ \mathrm{Re} \left\{ a b^\ast \left[
  \sqrt{5} \left( c^\ast d + g^\ast h \right)
  + 2 \sqrt{2} \left( d^\ast e + f^\ast g \right)
  + 3 e^\ast f \right] \right\},
\\
F_5 &=& \frac{2 \left( \left| \sqrt{5}\, c e - \sqrt{2}\, d^2 \right|^2
  + \left| \sqrt{5}\, f h - \sqrt{2}\, g^2 \right|^2 \right)}{9}
  + \frac{\left| \sqrt{5}\, c f - d e \right|^2
  + \left| \sqrt{5}\,  e h - f g \right|^2}{3}
  \\ & &
  + \frac{2 \left(\left| \sqrt{10}\, c g + d f - \sqrt{2}\, e^2 \right|^2
  + \left| \sqrt{10}\, d h + e g - \sqrt{2}\, f^2 \right|^2 \right)}{15}
  \\ & &
  + \frac{\left| 5 c h + 7 d g - 4 e f \right|^2}{45},
\\
F_6 &=& \frac{\left| 2 \sqrt{5}\, c g - 4 \sqrt{2}\, d f + 3 e^2 \right|^2
  + \left| 2 \sqrt{5}\, d h - 4 \sqrt{2}\, e g + 3 f^2 \right|^2
  + 2 \left| 5 c h - 3 d g + e f \right|^2}{35}.
\hspace*{12mm}
\ea
\es

\setcounter{equation}{0}
\renewcommand{\theequation}{B\arabic{equation}}

\section{Ans\"atze for the fields}
\label{app:phaseSpace}

In this appendix, we consider \textit{Ans\"atze} for the fields
in the cases with $n = 5$ and $n = 6$.

\subsection{$n = 5$}
\label{app:phaseSpace5}

With reference to Eqs.~\eqref{mvfp5} and~\eqref{A17},
we construct the following three Ans\"atze:
\begin{enumerate}
\item $d = e = g = 0$:
  If only $c$ and $f$ are nonzero,
  then
  \bs
  \ba
  \gamma_5 &=& \frac{3 F \left( 4 C + F \right)}{7 \left( C + F \right)^2},
  \\
  \gamma_6 &=& 0.
  \label{side1}
  \ea
  \es
  By letting $F$ vary from $0$ to $2 C$ one obtains
  $\gamma_5 \in \left[ 0,\ 4/7 \right]$.
\item $d = e = f = 0$:
  If only $c$ and $g$ are nonzero,
  then
  \bs
  \ba
  \gamma_5 &=& \frac{8 C G}{7 \left( C + G \right)^2},
  \\
  \gamma_6 &=& \frac{4 C G}{5 \left( C + G \right)^2}.
  \ea
  \es
  It is obvious that in this case
  \be
  \gamma_5 = \frac{10}{7}\, \gamma_6,
  \label{side2}
  \ee
  with $\gamma_5 \in \left[ 0,\ 2/7 \right]$
  and $\gamma_6 \in \left[ 0,\ 1/5 \right]$.
\item $d = f = 0$ and $g = - c$:
  In this case
  \bs
  \ba
  F_2 &=& 2 C + E,
  \\
  F_5 &=& \frac{16 C E}{7} + \frac{2 \left| 2 c^2 + e^2 \right|^2}{7},
  \\
  F_6 &=& \frac{\left| 2 c^2 - e^2 \right|^2}{5}.
  \ea
  \es
  Assuming furthermore that $c^2 = k e^2$,
  with a \emph{real} non-negative $k$,
  one obtains
  \bs
  \ba
  \gamma_5 &=& \frac{2}{7}
  \left[ 1 + \frac{8 k}{\left( 2 k + 1 \right)^2} \right],
  \\
  \gamma_6 &=& \frac{1}{5} \left( \frac{2 k - 1}{2 k + 1} \right)^2.
  \ea
  \es
  This gives
  \be
  \gamma_5 = - \frac{10}{7}\, \gamma_6 + \frac{4}{7},
  \label{side3}
  \ee
  with $\gamma_6 \in \left[ 0,\ 1/5 \right]$.
\end{enumerate}

\subsection{$n = 6$}
\label{app:phaseSpace6}

With reference to Eqs.~\eqref{mvfp6} and~\eqref{A20},
we construct the following five Ans\"atze:
\begin{enumerate}
\item $e = f = g = h = 0$: In this case
  \bs
  \ba
  \gamma_5 &=& \frac{4}{9} \left( \frac{D}{C + D} \right)^2,
  \\
  \gamma_6 &=& 0.
  \label{943oo}
  \ea
  \es
  The extreme situations $d=0$ and $c=0$ show that $0 \le \gamma_5 \le 4/9$.
\item $d = e = f = h = 0$: In this case
  \bs
  \ba
  \gamma_6 &=& \frac{4 C G}{7 \left( C + G \right)^2},
  \label{mvfppd}
  \\
  \gamma_5 &=& \frac{2 \left( 1 + 7 \gamma_6
    + \sqrt{1 - 7 \gamma_6} \right)}{9}.
  \label{3nvjfod}
  \ea
  \es
  Equation~\eqref{mvfppd} suggests that $0 \le \gamma_6 \le 1/7$,
  but in reality the curve~\eqref{3nvjfod} forms the boundary of the
  allowed region only for $0 \le \gamma_6 \le \varrho$,
  where
  \be
  \varrho = \frac{5 \left( 1361 + 288 \sqrt{5} \right)}{7 \times 109^2}
  \approx 0.121
  \ee
  is slightly less than $1/7 \approx 0.143$.
  The value of $\gamma_5$ corresponding
  through Eq.~\eqref{3nvjfod} to $\gamma_6 = \varrho$ is
  \be
  \varsigma = \frac{20 \left( 1607 + 471 \sqrt{5} \right)}{9 \times 109^2}
  \approx 0.498.
  \ee
\item $d = f = h = 0$: In this case
  \bs
  \ba
  F_2 &=& C + E + G,
  \\
  F_5 &=& \frac{10 C E + 4 G^2}{9}
  + \frac{4 \left| \sqrt{5}\, c g - e^2 \right|^2 + 2 E G }{15},
  \\
  F_6 &=& \frac{\left| 2 \sqrt{5}\, c g + 3 e^2 \right|^2
  + 32 EG}{35}.
  \ea
  \es
  Assuming furthermore that $e = \ell c$ and $g = - \sqrt{5}\, t c$,
  with \emph{real} $\ell$ and $t$,
  one obtains
  \bs
  \label{k_and_t}
  \ba
  \gamma_5 &=&
  \frac{50 \ell^2 + 500 t^4 + 300 t^2 + 12 \ell^4 + 120 \ell^2 t + 30 \ell^2 t^2}{45
    \left( 1 + \ell^2 + 5 t^2 \right)^2},
  \\
  \gamma_6 &=&
  \frac{100 t^2 + 9 \ell^4 - 60 \ell^2 t + 160 \ell^2 t^2}{35
    \left( 1 + \ell^2 + 5 t^2 \right)^2}.
  \ea
  \es
  We are interested in the envelope of this family of parametric curves.
  The relevant solution to the envelope equation reads
  \be
  \frac{\partial \gamma_6}{\partial t}\,
  \frac{\partial \gamma_5}{\partial \ell^2}
  - \frac{\partial \gamma_6}{\partial \ell^2}\,
  \frac{\partial \gamma_5}{\partial t}
  = 0
  \ \Leftrightarrow \
  \ell^2 = \frac{10 t \left( - 11 t^2 + 6 t + 1 \right)}{15 t^2 + 2 t +3}.
  \ee
  Plugging back this value of $\ell^2$ into Eqs.~\eqref{k_and_t}, 
  we obtain the parametric curve
  \bs
  \label{parametric}
  \ba
  \gamma_5 &=& \frac{20 t \left( 1125 t^7 - 195 t^6 + 821 t^5 - 1371 t^4
    + 823 t^3 + 399 t^2 + 175 t + 15 \right)}{9 \left(75 t^4
    - 100 t^3 + 90 t^2 + 12 t + 3 \right)^2},
  \\
  \gamma_6 &=& \frac{320 t^3 \left( - 165 t^4 + 212 t^3 - 102 t^2
    + 36 t + 3\right)}{7 \left( 75 t^4 - 100 t^3 + 90 t^2 +12 t + 3\right)^2}.
  \ea
  \es
  The parametric curve~\eqref{parametric}
  intersects the curves~\eqref{3nvjfod} and~\eqref{jgf000r},
  respectively,
  when
  \bs
  \ba
  \gamma_6 = \varrho &\Leftrightarrow&
  t=\frac{3 + 2 \sqrt{5}}{11},
  \\
  \gamma_6 = \frac{9}{70} &\Leftrightarrow& t = \frac{3}{5}.
  \ea
  \es
  In the interval $t \in \left[ 3/5,\,
  \left. \left( 3 + 2\, \sqrt{5} \right) \right/ 11 \right]$,
  the parametric curve~\eqref{parametric} may be approximated,
  to a high degree of accuracy,
  by the straight line\footnote{This 
  is the straight line connecting the points
  $\left( \varrho, \varsigma \right)$ and $\left( 9/70, 89/180 \right)$
  in the $\left( \gamma_6, \gamma_5 \right)$ plane.}
  \be
  \gamma_5 = \frac{21 \left( 71\,677 - 38\,000\, \sqrt{5}
    \right) \gamma_6 + 25 \left( 6\,485 + 4\,104\, \sqrt{5} \right)}{719\,298}.
    \label{parametric_approx}
  \ee
\item $e = f = 0$, $c = h$, and $d = - g$: In this case
  \bs
  \ba
  \gamma_6 &=& \frac{25 C^2 + 9 D^2 + 50 C D}{70 \left( C + D \right)^2},
  \label{nf9943} \\
  \gamma_5 &=& - \frac{14}{9}\, \gamma_6 + \frac{25}{36}.
  \label{jgf000r}
  \ea
  \es
  The straight line~\eqref{jgf000r} extends,
  according to Eq.~\eqref{nf9943},
  from $\gamma_6 = 9/70$ to $\gamma_6 = 5/14$.
  Note that when $\gamma_5$ and $\gamma_6$
  are related through Eq.~\eqref{jgf000r},
  $\delta$ is zero by virtue of Eq.~\eqref{22}.
\item $d = e = f = g = 0$: In this case
  \bs
  \ba
  \gamma_5 &=& \frac{5\, C H}{9 \left( C + H \right)^2},
  \\
  \gamma_6 &=& \frac{10\, C H}{7 \left( C + H \right)^2}.
  \ea
  \es
  It is clear that
  \be
  \gamma_5 = \frac{7}{18}\, \gamma_6,
  \label{prop}
  \ee
  with $5/14 \ge \gamma_6 \ge 0$ and $5/36 \ge \gamma_5 \ge 0$.
\end{enumerate}

\setcounter{equation}{0}
\renewcommand{\theequation}{C\arabic{equation}}

\section{Minimization of a function}
\label{app:math}

Consider the $\mathbb{R} \to \mathbb{R}$ function
\be
f(\theta) = \lambda_3 - k\left| \lambda_4 \right| \sqrt{p - s \theta}
+ \sqrt{\lambda_1}\, \sqrt{v + w \theta},
\ee
with real parameters $\lambda_1$,
$\lambda_3$,
$\lambda_4$, 
$k$,
$p$,
$s$,
$v$,
and $w$ satisfying
\begin{equation}
\lambda_1\geq0,\quad k\geq 0,\quad p>0,\quad s>0.
\label{assump}
\end{equation}
We are interested in the necessary and sufficient conditions that
guarantee $f(\theta)$ to be non-negative everywhere
inside the range $\theta \in \left[ \alpha, \beta \right]$,
with $\alpha \in \left[ 0, \beta \right]$ 
and $\beta \in\left[ 0, p/s \right]$.
It is \emph{necessary} that $f(\theta)$ be non-negative
at the end-points of the domain,
so the following two conditions must hold:
\bs
\label{necessary}
\ba
\lambda_3 - k\left| \lambda_4 \right| \sqrt{p - s \alpha}
+ \sqrt{\lambda_1}\, \sqrt{v + w \alpha} &\geq& 0,\\
\lambda_3 - k\left| \lambda_4 \right| \sqrt{p - s \beta}
+ \sqrt{\lambda_1}\, \sqrt{v + w \beta} &\geq& 0.
\ea
\es
If $k=0$ then $f(\theta)$ is monotonic,
so conditions~\eqref{necessary} are both necessary and sufficient.
If $k>0$ then $f(\theta)$ has at most one extremum:
\be
\left. \frac{\mathrm{d} f}{\mathrm{d} \theta} \right|_{\theta=\mu} =
\frac{k s |\lambda_4|}{2\sqrt{p - s\mu}}
+ \frac{w \sqrt{\lambda_1}}{2 \sqrt{v + w \mu}} = 0
\ \Rightarrow \
\mu = \frac{p w^2 \lambda_1 - k^2 s^2 v \lambda_4^2} {s w
  \left( w \lambda_1 + k^2 s \lambda_4^2\right)}.
\label{mut}
\ee
If that extremum is a minimum then we must avoid the situation 
where $\mu \in \left[ \alpha, \beta \right]$ and $f (\mu)$ is negative.
Therefore,
besides enforcing conditions~\eqref{necessary},
it suffices to avoid any situation where there is a $\mu$ such that
\bs
\label{ttt}
\ba
\left.\frac{\mathrm{d} f}{\mathrm{d} \theta}\right|_{\theta=\mu} &=& 0, \label{B4a}
\\
\left.\frac{\mathrm{d} f}{\mathrm{d} \theta}\right|_{\theta=\alpha} &<& 0, \label{B4b}
\\
\left.\frac{\mathrm{d} f}{\mathrm{d} \theta}\right|_{\theta=\beta} &>& 0, \label{B4c}
\\
f(\mu) &<& 0. \label{B4d}
\ea
\es
We aim to express conditions~\eqref{ttt} in terms of $\lambda_1$,
$\lambda_3$,
$\lambda_4$, 
$k$,
$p$,
$s$,
$v$,
and $w$.
To analyze condition~\eqref{B4a},
we observe that,
since $k s > 0$,
$\mathrm{d} f \left/ \mathrm{d} \theta \right.$ can only vanish if
\be
\label{negatived}
w < 0.
\ee
In this case,
the solution $\mu$ given in Eq.~\eqref{mut} satisfies
\bs
\ba
p - s \mu &=& \frac{k^2 s \lambda_4^2 \left( p w + s v \right)}{w
  \left( w \lambda_1 + k^2 s \lambda_4^2\right)},
\\
v + w \mu &=& \frac{w \lambda_1 \left( p w + s v \right)}{s
  \left( w \lambda_1 + k^2 s \lambda_4^2\right)}.
\ea
\es
Since $k > 0$,
$s > 0$,
and $w < 0$,
both $p - s \mu > 0$ and $v + w \mu > 0$ require
\be
\frac{p w + s v}{w \lambda_1 + k^2 s \lambda_4^2} < 0.
\label{93ofooc}
\ee
One then obtains
\bs
\ba
\sqrt{p - s \mu} &=& - \frac{k \sqrt{s} \left| \lambda_4 \right|}{w}\,
\sqrt{\frac{w \left( p w + s v \right)}{w \lambda_1 + k^2 s \lambda_4^2}},
\\
\sqrt{v + w \mu} &=& \sqrt{\frac{\lambda_1}{s}}\,
\sqrt{\frac{w \left( p w + s v \right)}{w \lambda_1 + k^2 s \lambda_4^2}},
\\
\left. \frac{\mathrm{d}^2 f}{\mathrm{d} \theta^2} \right|_{\theta = \mu} &=&
- \frac{w^2 \sqrt{s}}{4 k^2 \lambda_1 \lambda_4^2}
\left[ \sqrt{\frac{w \lambda_1 + k^2 s \lambda_4^2}{w
      \left( p w + s v \right)}}
  \right]^3
\left( w \lambda_1 + k^2 s \lambda_4^2 \right).
\ea
\es
Therefore,
in order that $\mu$ be a \emph{minimum} of $f(\theta)$ one must have
\bs
\label{B9}
\ba
w \lambda_1 + k^2 s \lambda_4^2 &<& 0, \label{B9a}
\\
p w + s v &>& 0,
\ea
\es
which is stronger than condition~\eqref{93ofooc}.
Notice that condition~\eqref{B9a} actually implies condition~\eqref{negatived},
so the latter is not needed any more.

Conditions~\eqref{B4b} and~\eqref{B4c} require $\mu$
to lie within the domain $\left[ \alpha, \beta \right]$.
Those conditions translate into
\bs
\label{B10}
\ba
\frac{k s |\lambda_4|}{\sqrt{p - s \alpha}} +
\frac{w \sqrt{\lambda_1}}{\sqrt{v + w \alpha}} &<&  0, \label{B10a}
\\
\frac{k s |\lambda_4|}{\sqrt{p - s \beta}} +
\frac{w \sqrt{\lambda_1}}{\sqrt{v + w \beta}} &>&  0. \label{B10b}
\ea
\es
It is readily seen that
condition~\eqref{B10a} is equivalent to $\mu > \alpha$,
and that condition~\eqref{B10b} is equivalent to $\mu < \beta$,
when $\mu$ is given by Eq.~\eqref{mut}.

Finally,
$f(\mu)$ is negative when
\ba
\label{B11}
\lambda_3 + \sqrt{\frac{\left(p w + s v \right)
\left(w \lambda_1 + k^2 s \lambda_4^2  \right)}{s w}}
< 0.
\ea

To summarize,
the \emph{necessary and sufficient} conditions
for $f(\theta)$ to be non-negative
in the whole interval $\left[ \alpha, \beta \right]$ are
conditions~\eqref{necessary} if $k=0$.
If $k>0$,
then one must furthermore \emph{exclude}
the situation where all five inequalities~\eqref{B9},
\eqref{B10},
and~\eqref{B11} simultaneously hold.

\end{appendix}

\end{document}